\documentclass[12pt,english,showpacs, amsfonts,amsmath]{revtex4}

\usepackage{graphicx}
\usepackage{epstopdf}
\usepackage{amssymb}
\usepackage{amsmath}
\usepackage{hyperref}
\usepackage{xspace}
\usepackage{color}

\newcommand{\bra}[1]{\left\langle #1 \right|}
\newcommand{\ket}[1]{\left| #1 \right\rangle}
\newcommand{\aket}[1]{\left| \phi_{#1} \right\rangle}
\newcommand{\abra}[1]{\left\langle \phi_{#1} \right|}

\begin{document}

\title{Quantum Correlations in Cavity QED Networks}

\author{Miguel Orszag$^{1,*}$, Nellu Ciobanu$^{1}$, Raul Coto$^{1}$, Vitalie Eremeev$^{2}$ }
\affiliation
{$^{1}$Facultad de F\'{i}sica, Pontificia Universidad Cat\'{o}lica de Chile, Casilla 306, Santiago, Chile\\
$^{2}$Facultad de Ingenier\'{i}a, Universidad Diego Portales, Santiago, Chile}

\email{* Email:morszag@fis.puc.cl}

\begin{abstract}
We present a Review of the dynamical features such as generation, propagation, distribution, sudden transition and freezing of the various quantum correlation measures, as Concurrence, Entanglement of Formation, Quantum Discord, as well their geometrical measure counterparts within the models of Cavity Quantum Electrodynamics Networks.
Dissipation and thermal effects are discussed both in the generation of quantum correlations as well as their effect on the sudden changes and freezing of the classical and quantum correlations in a cavity quantum electrodynamical network.
For certain initial conditions, double transitions in the Bures geometrical discord are found. One of these transitions tends to disappear at a critical temperature.

\end{abstract}

 \maketitle

\section{Introduction}

Quantum correlations play a fundamental role in quantum computation and quantum information processing  \cite{Nielsen, Modi12, Horodecki}, where entanglement is usually considered a popular measure of such correlations. Entanglement (verschr\"ankung) introduced in physics originally by Schr\"odinger \cite{Schrodinger} and considered a native feature of the quantum world, is the most outstanding and studied phenomenon to test the fundamentals of quantum mechanics, as well as an essential engineering tool for the quantum communications. However,
 entanglement is a property that is hard to reach technologically and even when achieved, it is a very unstable quantum state,  vulnerable
 under the effects of decoherence, any dissipative process as a result of the coupling to environment. Conventionally these effects
 are considered mainly destructive for entanglement, nevertheless some recent studies of this subject attest results different from the
 common conviction, even appearing as counterintuitive at first glance \cite{Cirac11, Sorensen, Memarzadeh}.

An alternative approach to measure the entire correlations in a quantum system was suggested originally in Refs. \cite{Vedral, Zurek}. By using the concepts of mutual information and quantum discord (QD) the quantum correlations may be distinguished from the classical ones.
Further, the QD could be compared to the entanglement of formation (E) \cite{Wootters} or relative entropy of entanglement (REE) \cite{Vedral97} in order to find out if the system is in a quantum inseparable state (entangled), or in a separable state with quantum correlations, such as QD \cite{Luo, Alber, Lu, Fanchini}. In the last few years, some alternative measures of the QD were proposed and studied intensively. These measures are related basically to the entropic and geometric quantities of QD-like correlations, for an exhaustive review see Ref. \cite {Modi12}, where the most common non-classical correlations are: quantum dissonance proposed in \cite{Modi10}; geometric QD (GQD) based on the trace norm as Hilbert-Schmidt distance \cite{Dakic10}, Schatten p-norm \cite{Debara} and Bures distance \cite{Spehner, Aaronson2013}, \textit{etc.}. The QD has shown to capture non-classical correlations including the completely separable systems, e.g., deterministic quantum computation with one quantum bit (DQC1) model \cite{Datta08}.

In this Review we propose to analyze broadly the phenomena such as generation, propagation, distribution and measurement of quantum and classical correlations in a particular kind of quantum systems known as cavity quantum electrodynamics (CQED) networks which deal with atoms placed in cavities interconnected by fibers, in the framework of the
physical model suggested in Ref. \cite{Cirac97}, which attracted a high interest for quantum information applications and subsequently
developed from different aspects \cite{Pellizzari, Mancini, Serafini, Zheng, Ritter}. The inclusion of the interaction of the quantum system (atoms + fields) with the environment plays an important role in physics, implying a more realistic picture because the dissipation is always present in the real devices. The entire system is considered open because of the leakage of the electromagnetic field from the cavities and fiber into their own reservoirs. We initiated our investigations few years ago by proposing a 2-node CQED network: two atoms (qubits) each one trapped in a cavity and interconnected by fiber, with coupling of each quantum subsystem to the individual Markovian thermal environment, see Fig.1 in \cite{Mont}. In this system, in the approximation of one excitation, for all the reservoirs at zero temperature we found the effect of generation and oscillation in time of the entanglement measured by the concurrence as shown in Figs. 3-4 in \cite{Mont} for an initially separable  state and sharing one excitation between the two qubits. Also, in the same work, the possibility of preservation of the entanglement at its maximal value for a period of time even in the presence of losses was proposed, by managing the atom-cavity detuning as shown in Figs. 5-8 of \cite{Mont}. Next we investigated  the influence of the reservoirs' temperature on the classical and quantum correlations. For example, in the paper \cite{MEO} we have shown that for the initial states similar to the ones considered in \cite{Mont} it is possible to stimulate the enhancement of the maximal entanglement by the thermal reservoirs up to a particular "critical" temperature, beyond which the entanglement starts decreasing, as shown in Fig. 3 in \cite{MEO}. Inspired by this effect of entanglement gain by the thermal environments in the given CQED network, we developed our study further. So, in \cite{EMO} we demonstrated that it is possible, for a two-qubit system, initially in zero-excitation state, to generate  long-lived quantum correlations as entanglement and quantum discord with the assistance of thermal environments, see Figs. 2-5 in \cite{EMO}, and the optimal situation is found using the fiber thermal reservoir, as shown in Fig. 6, ibid. Hence, we came to the conclusion that it is possible to generate atomic quantum correlations in CQED networks with dissipation channels by the processes of absorption and emission, i.e. exchanging excitations with the thermal reservoirs. 

\begin{figure}[ht]
\centering
\includegraphics[width=0.8 \textwidth]{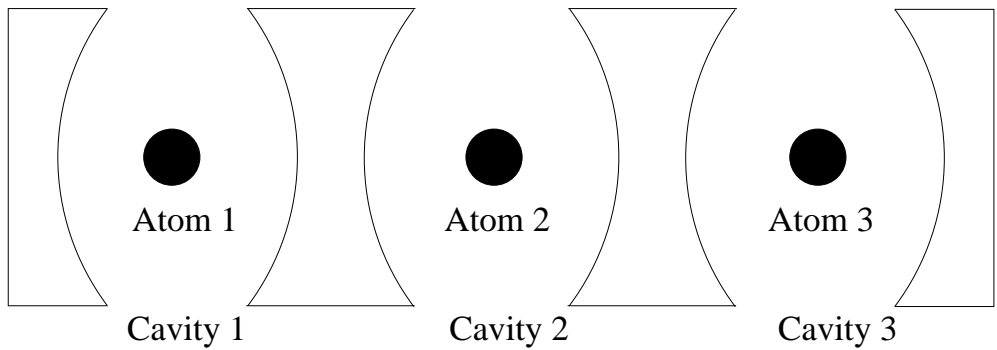}
\caption{Array of three Cavity-Atom Systems.}
\label{fig1}
\end{figure}

The propagation of quantum correlations, over the past decades, have captured the attention of many researchers due to it`s powerful applications in a wide range of physics \cite {Nielsen}. Cavity QED networks, are particularly convenient for the creation and propagation of these correlations.
There are several ways to build a system for quantum computation or communication, depending on the distribution of the cavities, the way these are coupled together, boundary conditions, etc. The most typical is a chain of cavities, see (Fig. \ref{fig1}). 

We can in principle, couple neighboring cavities in two ways, either via an optical fiber \cite{Pellizzari, Mancini, Serafini, Mont, EMO, zhong, Zhang, mancini} or by tunnel effect \cite{cavities,angelakis1,angelakis2}. In the latter, the cavities need to be close enough so that photon hopping can occur due to the overlap of the spatial profile of the cavity modes. The former type of coupling becomes important mainly when large distance needs to be covered, e.g., quantum communication between two distant nodes in the network, for experiment see \cite{Ritter}. Also, it may be useful in generating photon phases when going from one cavity to the other \cite{Serafini}. 

Multipartite Quantum Correlations is one of the most relevant indicator of the quantumness of a system in many body systems. This remarkable feature is in general difficult to characterize and the known definitions are  hard to measure. Besides the efforts dedicated to solve this problem, the question of which is the best approach remains open. 

Many attempts of extending the bipartite correlations to the multipartite case have been made \cite{wootters, multipostulates,woottersGeneral,coefficientmatrix,zambrini,fanchini,witness,sarandy}, but still questions remain about these generalizations. The first approach was the \textit{Tangle} \cite{wootters}, which is related to the entanglement. In that paper, the authors argue that unlike classical correlations, quantum entanglement cannot be freely shared among many objects. For example, in the case of three partite system, labeled as ``$1$", ``$2$" and `$3$"; the amount of entanglement that the first system can share with the two others, must satisfy the inequality:
 
\begin{equation}\label{relation}
\mathit{C}_{12}^2 + \mathit{C}_{13}^2\leq 4\det[\rho_1]
\end{equation}
with $\rho_1=tr_{23}[\rho_{123}]$. The above equation can be rewritten as $\mathit{C}_{12}^2 + \mathit{C}_{13}^2\leq \mathit{C}_{1(23)}^2$, for the case of pure states. Then, it is defined the quantity,
\begin{equation}\label{tangle}
\tau_{123}=\mathit{C}_{1(23)}^2-\mathit{C}_{12}^2 - \mathit{C}_{13}^2
\end{equation}

This residual entanglement represents a collective property of the three qubits that is unchanged by permutations; it
measures an essential three-qubit entanglement. In words, the entanglement of ``$1$" with ``$23$" can be manifested
in three forms, entanglement with ``$2$", entanglement with ``$3$", and an essential three-way entanglement.
These three forms must share the total entanglement. 

For the case of mixed states $\rho$, $\mathit{C}_{1(23)}(\rho)$  is not defined. However, one can consider all possible pure-state decompositions of the state $\rho$, that is, all sets $\lbrace (\psi_i,p_i)\rbrace$ such that $\rho=\sum_i p_i\vert \psi_i\rangle\langle\psi_i\vert$. For each of these decompositions, one can compute the average value $\langle\mathit{C}_{1(23)}^2\rangle(\rho)=\sum_i p_i\mathit{C}_{1(23)}^2(\psi_i)$. Then, with the minimum of this average over all decompositions of $\rho$, the analogue of Eq.(\ref{tangle}) for mixed state will be;

\begin{equation}\label{tanglemix}
\tau_{123}(\rho)=(\mathit{C}_{1(23)}^2)^{min}-\mathit{C}_{12}^2 - \mathit{C}_{13}^2
\end{equation}

This task is usually computational expensive, but there are some good approximations \cite{upperbound,lowerbound,quasipure,Davidovich}. The first two references correspond to an upper and lower bound respectively. The main idea is to narrow down the values of the tangle with these bounds to get closer to the real value. Before explaining both methods, it is important notice that the term in the right side of Eq.(\ref{relation}) can be rewritten such that $\mathit{C}_{1(23)}^2=2(1-tr[\rho_1^2])$ \cite{buzek}. This form is more convenient as we will see next. The upper bound is found by taking just a pure state, which means using Eq.(\ref{tangle}) as if the system was pure. The lower bound comes from computing  $\mathit{C}_{1(23)}^2(\rho)=2(tr[\rho^2]-tr[\rho_1^2])$, where $tr[\rho^2]$ is the purity of the total system. 

In the reference \cite{quasipure}, the authors did a quasi-pure approximation, but their procedure is not as simple as the one used in \cite{upperbound,lowerbound}. In order to find an exact solution, a conservation law for the distributed entanglement of formation and quantum discord has been found \cite{winter,caldeira}. This method, which works fine only for small dimensional systems, it is well explained in reference \cite{fanchini}.

Another kind of  interesting effects observed in  quantum open systems are related to the unusual dynamics of the classical and quantum decoherence originally reported in \cite{Maziero2009, Mazzola2010} and confirmed experimentally in \cite{Xu, Auccaise2011}, hence stimulating a high interest in the investigation of the phenomena of sudden changes in the correlations for different physical systems. During the last years, intensive efforts have focused to explain the nature of the sudden transitions and freezing effects of the quantum correlations and the conditions under which such transitions occur. Also, from the perspective of the applications, how efficiently one could engineer these phenomena in quantum technologies. As has been shown in the studies \cite{Maziero2009, Auccaise2011, Xu, Pinto2013, He2013, Mazzola2010, Lo Franco2012, You2012, Aaronson2013}, the puzzling peculiarities of the sudden transitions and freezing phenomena are hidden in the structure of the density operator during the whole evolution of a bipartite quantum system for particular decoherence processes. Nevertheless, important questions remain open - how these fascinating effects are affected by the presence of the noisy environments and if there are efficient mechanisms to control them in both non dissipative or dissipative decoherence models. The state-of-the-art research of CQED networks has shown so far a modest progress on the influence of the environments on the the sudden transitions and freezing phenomena \cite{He2013}, with very little research on the influence of thermal baths in such quantum open systems. Motivated by this interest in the field, very recently in \cite{ECO} we presented some novel results concerning the sudden transitions and freezing effects of the quantum correlations for the same CQED network as in Fig.  \ref{fig1ol}, but developed for many excitations in the whole system and including the environments. We have shown that the detrimental effect of the thermal reservoirs on the freezing of correlations can be compensated via an efficient coupling of the fiber connecting the two cavities. Furthermore, for certain initial conditions, a double sudden transition in the dynamics of the Bures geometrical quantum discord was observed. Similar double transitions were reported for Schatten one-norm geometric quantum correlations (GQD-1) in \cite {Montealegre2013, Paula2013}. In our system, the second transition tends to disappear at a critical temperature, hence freezing the discord. We concluded that by controlling the dissipation mechanisms it is possible to engineer  sudden changes and freezing periods in the temporal evolution of the quantum correlations with multiples effects which can find practical applications. This kind of critical thermal effects appear in CQED networks as well as other  systems \cite {Werlang, Hu}

\section{Classical and Quantum Correlations}

\textbf{Entanglement of formation} 
\newline For a given ensemble of pure states $\left\{ p_{i},\mid \psi _{i}\rangle \right\} $, the entanglement
of formation is the average entropy of entanglement over a set of states that minimizes this average over all possible decompositions of $\rho$,
\cite{BENNETT96}.
\begin{equation}
E(\rho )=\min{\sum_{i}p_{i}E(\psi _{i})},
\end{equation}
where the entanglement $E(\psi )$ is defined as the von Neumann entropy of
either one of the  subsystems $E(\psi _{A/B} )=S(\rho _{A/B})$, with $%
S(\rho )=-tr[\rho \log _{2}\rho ]$. However, it is very difficult to know which ensemble $\{p_{i},\psi _{i}\}$
is the one that minimizes the average. A concept closely related to the
entanglement of formation is the concurrence \cite{Wootters, WOOTTERS97}.
\newline
For a general mixed state $\rho _{AB}$ of two qubits, we define $\widetilde{%
\rho }$ to be the spin-flipped state $\widetilde{\rho }_{AB}=(\sigma
_{y}\otimes \sigma _{y})\rho _{AB}^{\ast }(\sigma _{y}\otimes \sigma _{y})$
where $\rho ^{\ast }$ is the complex conjugate of $\rho $ and $\sigma _{y}$
is the Pauli matrix. The concurrence is defined as
\begin{equation}
C(\rho )=\text{max}\{0,\lambda _{1}-\lambda _{2}-\lambda _{3}-\lambda
_{4}\},
\end{equation}
where $\{\lambda _{i}\}$ are the square roots in decreasing order of the
eigenvalues of the non-hermitian matrix $\rho \widetilde{\rho }$. \newline
Finally, the entanglement of formation is related to concurrence as follows
\begin{equation}
E(\rho )=H \left( \frac{1}{2}+\frac{1}{2}\sqrt{1-C^2} \right)
\label{enta}
\end{equation}
with $H(x)=-x\log _{2}x-(1-x)\log _{2}(1-x)$. \newline The entanglement vanishes for a \emph{separable} state, defined as
\begin{equation}
\rho =\sum_{\scriptscriptstyle{i}}p_{\scriptscriptstyle{i}}{\rho _{\scriptscriptstyle{i}}}^{\scriptscriptstyle{A}}\otimes {\rho _{\scriptscriptstyle{i}}}^{\scriptscriptstyle{B}}
\end{equation}
and it is equal to one for maximally entangled states.

\textbf{Quantum Discord} 
The total correlations of a quantum system are
quantified by the quantum mutual information $I(\rho )=S(\rho ^{A})+S(\rho ^{B})-S(\rho )$. The total amount of correlations can be separated into
classical and quantum correlations $I(\rho )=C(\rho )+Q(\rho )$. In search of a formula for measuring the classical correlations, Henderson and Vedral proposed a list
of conditions that the measure of classical correlations must satisfy \cite{Vedral}. Correspondingly they proposed a quantifier that fulfilled all the conditions, so the classical correlations are measured as
\begin{equation}
C(\rho ^{AB})=\displaystyle\max_{\{B_{k}\}}{[S(\rho ^{A})-S(\rho^{AB} |\{B_{k}\})]%
}  \label{clasi}
\end{equation}%
with the quantum conditional entropy of A defined as $S(\rho^{AB} |\{B_{k}\})=%
\displaystyle\sum_k{p_{k}S(\rho _{k})}$, where $\{\rho _{k},p_{k}\}$
is the ensemble of all possible results for the outcome from the set of von
Neumann measurements $\{B_{k}\}$ made in the subsystem $B$. Also $\rho _{k}=(I\otimes B_{k})\rho (I\otimes B_{k})/p_{k}$ is the  state of the system after a measurement, where $p_{k}=tr(I\otimes B_{k})\rho (I\otimes
B_{k})$ is the probability for obtaining the outcome $k$ after the measurement. The maximization in Eq(\ref{clasi}) is done over all possible
measurements of B, which implies to look for the measurement that disturbs the least the overall quantum state. \newline With this definition for
the classical correlation, we get the Quantum Discord as $QD(\rho )=I(\rho )-C(\rho )$. For pure states, this formula coincides with entanglement
of formation. \newline A problem with the QD is that it is asymmetrical with respect to which part of the bi-partite system is measured. However
it becomes symmetrical for particular systems with $S(\rho ^{A})=S(\rho ^{B})$.
\newline 

\textbf{Geometric quantum discord (GQD) and geometric entanglement (GE)}

% During the last few years, there different measures have been proposed to quantify the GQD, for a review see \cite {Modi12}. Geometric QD based on the trace norm as Hilbert-Schmidt distance \cite{Dakic10}, Schatten p-norm \cite{Debara, Montealegre2013} and Bures distance \cite{Spehner, Aaronson2013}.

In this Review we will use the calculations of GQD for two-qubit states with maximally mixed marginals (Bell-diagonal) as in Eq.(\ref{rhoBD}) measured by the Bures distance, which is the minimal geometric distance of a quantum state of a bipartite system AB to the set of classical states for subsystem A \cite{Spehner}. Based on this reference, we present here the main formulas used in our computations of the Bures GQD quantified by the normalized quantity $\tilde{D}_A$ as follows

%Bures QD equation
\begin{equation}
\widetilde{D}_A(\rho) = \left(1-\frac{1} {\sqrt{2}}\right)^{-1} \left(1-\sqrt{\frac{1+b_{\text{max}}}{2}}\right)
\label{BuresQD}
\end{equation}
with 
\begin{eqnarray}
b_{\text{max}}=\frac{1}{2} \text{max} \big \{ \sqrt{(1+c_1)^2-(c_2-c_3)^2}+\sqrt{(1-c_1)^2-(c_2+c_3)^2}, \nonumber \\ 
\sqrt{(1+c_2)^2-(c_1-c_3)^2}+\sqrt{(1-c_2)^2-(c_1+c_3)^2}, \nonumber \\
 \sqrt{(1+c_3)^2-(c_1-c_2)^2}+\sqrt{(1-c_3)^2-(c_1+c_2)^2} \big \},
 \label{bmax}
\end{eqnarray}
and the Bell-diagonal (BD) density matrix is defined as
%Bell-diagonal rho matrix
\begin{equation}
\rho_{BD}=[I\otimes I+\vec{c} \cdot (\vec{\sigma} \otimes \vec{\sigma})]/4=\frac{1}{4}\begin{pmatrix}  1+c_3 &0&0&c_1-c_2\\ 0& 1-c_3& c_1+c_2&0\\ 0& c_1+c_2& 1-c_3&0\\ c_1-c_2&0&0&1+c_3 \end{pmatrix},
 \label{rhoBD}
\end{equation}
where $\vec{\sigma}=(\sigma_1, \sigma_2, \sigma_3)$ is the vector given by Pauli matrices, $I$ is the identity matrix, and the vector $\vec{c}=(c_1, c_2, c_3)$ defines completely the state with $-1 \le c_i \le1$. 

Similarly to the entropic entanglement such as entanglement of formation (E) and relative entropy of entanglement (REE) \cite {Modi12, Wootters, Vedral97}, which are used often for comparison to the entropic QD, one can define the geometric measure of entanglement (GE), that for the two qubits case is given by \cite{Spehner}
%Geom entanglement GE
\begin{equation}
GE(\rho) = 2-\sqrt{2}\left(1+\sqrt{1-C(\rho)^2}\right)^{1/2},
 \label{GE}
\end{equation}
where $C(\rho)=\text{max} \{ \vert{c_1-c_2}\vert-1+c_3, \vert{c_1+c_2}\vert-1-c_3, 0 \}/2$ is the Wooters concurrence \cite{Wootters} computed here for the BD matrix. The normalized  geometric entanglement is $\widetilde{GE}(\rho) =GE(\rho)/(2-\sqrt{2})$ whose dynamics will be compared to Bures GQD, $\widetilde{D}_A(\rho)$, in the next section.

\section{Generation and criticallity of Correlations}
\subsection{2-node CQED network with dissipations to thermal reservoirs}
\begin{figure}[t]
\includegraphics[width=7 cm]{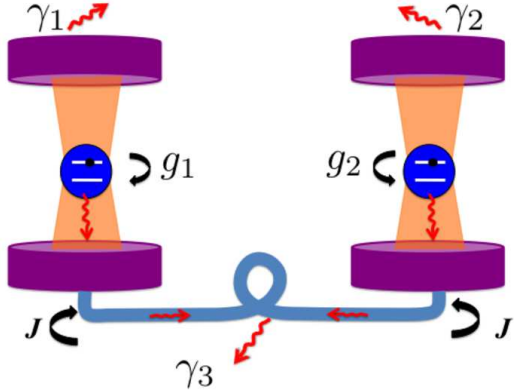}
\caption{Two atoms trapped in distant coupled cavities. The cavities and transmission line exchange the energy at the rates
  $\gamma_1$, $\gamma_2$ and $\gamma_3$ with their baths having the temperatures $T_1$, $T_2$ and $T_3$, respectively.}
  %%$g_i$ and J are the atom-cavity and cavity-fiber coupling constants.
\label{fig1ol}
\end{figure}

We present here the model schematically shown in Fig.\ref{fig1ol} where the two remote qubits (two-level atoms) interact with individual cavity and coupled
by a transmission line (e.g., fiber, waveguide). For simplicity we consider the short fiber limit: only one mode of the fiber
 interacts with the cavity modes \cite{Serafini}.
The Hamiltonian of the system under the rotating-wave approximation (RWA) in units of $\hbar$ reads
%\begin{eqnarray}
%system Hamiltonian
\begin{align}
H_s &= \omega_f a_3^{\dag} a_3+\sum\nolimits_{j=1}^2  \left( \omega_a S_{j,z}+\omega_0 a^{\dag}_j a_j \right) \nonumber \\
&+ \sum\nolimits_{j=1}^2 \left( g_j S^+_j a_j + J a_3 a^{\dag}_j + H.c.\right),
\label{Ham}
\end{align}
%\end{eqnarray}
where $a_1(a_2)$ and $a_3$ is the boson operator for the cavity 1(2) and the fiber mode, respectively;
 $\omega_0$, $\omega_f$  and $\omega_a$ are the cavity, fiber and atomic frequencies, respectively;  $g_j (J)$ the atom(fiber)-cavity coupling constants; $S_{z}$, $S^{\pm}$ are the atomic inversion and ladder operators, respectively.

%Fig.1a of OL_ECO 

One of the important advance in our proposal of CQED network model is based on the generalization to large number of excitations in the whole system. To the best of our knowledge, this approach of many excitations in similar systems \cite{Cirac97, Pellizzari, Mancini, Serafini, Zheng} is not common, and may be one of few existent studies. To describe the evolution of an open quantum-optical system usually the approach of the Kossakowski-Lindblad phenomenological master equation is considered with the system Hamiltonian decomposed on the eigenstates of the field-free subsystems. However, sometimes a CQED system is much more realistically modeled based on the microscopic master equation (MME), developed in \cite{Scala, Breuer} where the system-reservoir interactions are described by a master equation with the system Hamiltonian mapped on the atom-field eigenstates, known as dressed states. In our case the system consists in two atoms within their own cavities connected by a fiber and we consider the leakage of the two cavities and the fiber via a coupling to individual external environments, thus identifying three independent dissipation channels. Commonly, in CQED the main sources of dissipation originate from the leakage of the cavity photons due to the imperfect reflectivity of the cavity mirrors. Another mechanism of dissipation corresponds to the spontaneous emission of photons by the atom, however this kind of loss is negligible small in the CQED regime considered in our model, and consequently is neglected. Hence, it is straightforward to bring the Hamiltonian $H_s$ in Eq. (\ref{Ham}) to a matrix representation in the atom-field eigenstates basis. To define a general state of the whole system we use the notation: $\ket{i}=\ket{A_1}\otimes\ket{A_2}\otimes\ket{C_1}\otimes\ket{C_2}\otimes\ket{F} \equiv \ket{A_1A_2C_1C_2F}$, where 
${A}_{1,2}$ correspond to the atomic states, that can be $e(g)$ for excited(ground) state, while ${C}_{1,2}$ and ${F}$ define the cavities and fiber states, respectively, which may correspond to $0$, $1$, ..., $n$ photon states. Because the quantum system is dissipative, the excitations may leak to the reservoirs degrees of freedom, hence the ground state of the system, $\ket{0}=\ket{gg000}$, should be also considered in the basis of the states. Therefore, in the case of $N$ excitations in our system, the number of dressed states, $\ket{i}$, having minimum one excitation, i.e excluding the ground state $\ket{0}$, is computed by a simple relation: $d_N=N+2\sum_{k=1}^N k(k+1)$. For example, in case of $N=2$ excitations the Hamiltonian $H_s$ in Eq. (\ref{Ham}) is decomposed in a state-basis of the dimension $1+d_2$, i.e. is a  $19\times19$ matrix; for 6 excitations $H_s$ is represented  by a $231\times231$ matrix, and so on. Hence it is evident that for large $N$ the general problem becomes hard to solve even numerically. In our work \cite{Mont,MEO,EMO,ECO} we developed the calculations from the simplest case of $N=1$ up to 6 excitations, which is an improvement as compared to some previous works at similar subject, e.g. with two excitations \cite{Serafini}. In the present Review we present some results with $N=9$ excitations.

\begin{figure} [t]
\includegraphics[width=0.85 \textwidth]{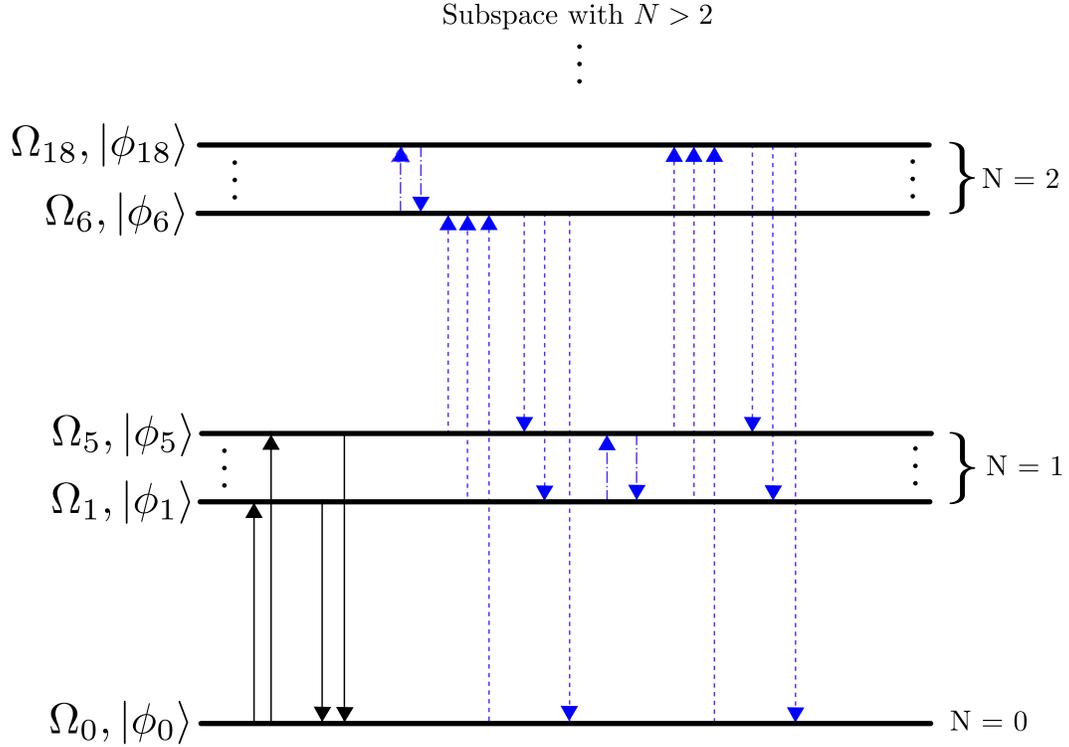}
\caption{The schematic representation of the transitions in the space of dressed states of the system Hamiltonian Eq. (\ref{Ham}) for N excitations.}
\label{figDS}
\end{figure}

Considering the above assumptions and following the approach of \cite{Scala, Breuer}, the MME for the reduced density operator $\rho(t)$ of the system is derived
%MME equation
\begin{equation}
\frac {\partial \rho}{\partial t}=-i\left[ H_s,\rho \right]+\mathcal{L}(\bar{\omega}) \rho+\mathcal{L}(-\bar{\omega}) \rho,
\label{MME}
\end{equation}
where the dissipation terms are defined as follows (with $\bar{\omega}>0$)
\begin{equation}
\mathcal{L}(\bar{\omega}) \rho = \sum_{j=1}^3 \gamma_j(\bar{\omega})\bigg(A_j(\bar{\omega})\rho A_j^ \dag (\bar{\omega})- \frac{1}{2} \left[ A_j^
\dag (\bar{\omega}) A_j(\bar{\omega}),\rho \right] _{+} \bigg). 
\label{Lind}
\end{equation}
In the above equations the following definitions are considered: $A_j(\bar{\omega}) = \sum_{\bar{\omega}_{\alpha, \beta}} \aket{\alpha}
\abra{\alpha} (a_j + a_j^{\dag}) \aket{\beta}\abra{\beta}$ fulfilling the properties $A_j(-\bar{\omega})= A_j^{\dag}(\bar{\omega})$,
where $\bar{\omega}_{\alpha, \beta} = \Omega_{\beta} - \Omega_{\alpha}$ with $\Omega_k$ as an eigenvalue of Hamiltonian $H_s$ and its corresponding eigenvector $\aket{k}$, denoting the \textit{k}-th dressed-state (see Fig.\ref{figDS}). 

%Fig of dressed-states transitions

In order to solve such a MME we will use the numerical simulations, because in the most general case there is almost impossible to find the analytic solution for the eigenvalue equation based on Hamiltonian $H_s$ (\ref{Ham}). In the following we develop the equation for the density operator $\rho(t)$ mapped on the eigenstates basis, $\bra{\phi_m}\rho(t)\ket{\phi_n} = \rho_{mn}$ for the case of $N$ excitations in the system
%rho_mn eq
\begin{eqnarray}
\dot{\rho}_{mn}= - i \bar{\omega}_{n,m} \rho_{mn} + \sum_{k=1}^{d_N} \big[ \frac{\gamma_{k \to 0}}{2} \big( 2\delta_{m0}\delta_{0n} \rho_{kk} - \delta_{mk}\rho_{kn}  - \delta_{kn} \rho_{mk} \big)\nonumber \\ 
+\frac{\gamma_{0 \to k}}{2} \big( 2\delta_{mk}\delta_{kn}  \rho_{00} - \delta_{m0}\rho_{0n} - \delta_{0n}\rho_{m0}  \big) \big],
 \label{rhosys}
\end{eqnarray}
here $\delta_{mn}$ is the Kronecker delta; the physical meaning of the damping coefficients $\gamma_{k \to 0}$ and $\gamma_{0 \to k}$ refer to the rates of the transitions between the eigenfrequencies $\Omega_k$ and $\Omega_0$ downward and upward, respectively, defined as follows $\gamma_{k \to 0}=\sum_{j=1}^3 c_i^2\gamma_j(\bar{\omega}_{0,k}) \left[\langle n(\bar{\omega}_{0,k})\rangle_{T_j} + 1\right] $ and by the Kubo-Martin-Schwinger (KMS) condition we have $\gamma_j(-\bar{\omega})=\mathrm{exp}\left(-\bar{\omega}/ T_j\right) \gamma_j({\bar{\omega}})$, where $c_i$ are the elements of the transformation matrix from the states  $\{\ket{0}, \ket{1}, ... , \ket{d_N}\}$ to the states $\{\aket{0}, \aket{1}, ... , \aket{d_N} \}$ (similar to Eq. (14) and Appendix A in \cite{Mont}). The KMS condition ensures that the system tends to a thermal equilibrium for $t \to \infty$. Here $\langle n(\bar{\omega}_{\alpha, \beta})\rangle_{T_j} = \left ( \mathrm{e}^{(\Omega_\beta- \Omega_\alpha) / T_j} - 1\right )^{-1}$ corresponds to the average number of the thermal photons (with $k_B=1$). The damping coefficients play a very important role in our model because their dependence on the reservoirs temperatures imply a complex exchange mechanism between the elements of the system and the baths. Further, one solves numerically the coupled system of the first-order differential equations (\ref{rhosys}) and compute the evolution of different kind of correlations between the two distant atoms, given some finite temperature of the reservoirs. In order to get the reduced density matrix for the atoms one performs a measurement on the cavities and the fiber vacuum states, $\ket{000} = \ket{0}_{C1}\otimes \ket{0}_{C2} \otimes \ket{0}_{F}$. Later, we will explain how this task can be realized experimentally. We find that, after the projection, the reduced atomic density matrix has a X-form and the quantum and classical correlations can be computed easily as developed in \cite{Luo, Alber, Fanchini, Spehner}. 

%Fig.2 of PRA_EMO
\begin{figure*} [h]
\includegraphics[width=0.95 \textwidth]{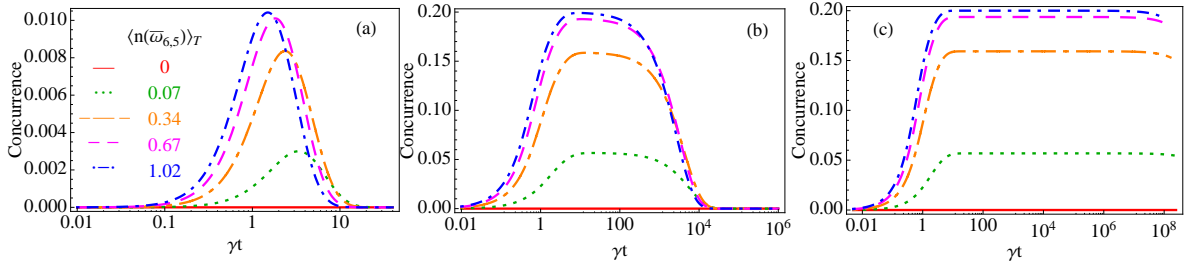}
\caption{Evolution of the concurrence for $g=J=5\gamma$ and different atom-cavity
detunings: (a) $\Delta =0$ , (b) $\Delta=10^{-4}\omega_a $ and (c) $\Delta=0.1 \omega_a $. The baths have the same temperature with the average number of thermal photons given by $\langle n(\bar{\omega}_{6, 5})\rangle_{T}$. The axis of the dimensionless time, $\gamma t$, is in a
logarithmic scale.}
\label{fig2pra}
\end{figure*}
In the following, we present the results obtained recently using this model. Beginning with the first explorations in \cite{Mont}, we  observed few interesting effects. Considering the baths at zero temperature and the initial two-qubit state as separable  sharing one excitation, the generation and oscillation in time of the entanglement was found. Additionally, the conservation of the maximal value of the entanglement for a period of time in the presence of losses was obtained with specific  atom-cavity detunings, as can be seen in Figs. 5-8 of \cite{Mont}. Inspired by these findings, we proceed further  \cite{EMO}  exploring the time evolution of the atomic entanglement measured via concurrence and entanglement of formation, as well as  the classical correlations and quantum discord, all these quantities as functions of the temperature of the reservoirs.

 The system under  numerical study in \cite{EMO} considered atoms with long radiative lifetimes, each coupled to its own cavity. The two cavities are connected by a fiber with the damping rates $\gamma_1=\gamma_2=\gamma_3 \equiv \gamma=2 \pi$ MHz, respectively, which are within the current technology \cite{Serafini}. The transition frequency of the atom is chosen to be mid-infrared (MIR), i.e. $\omega_a/2 \pi=4$THz and hence, for experimental purposes the coupling between the distant cavities can be realized by using the modern resources of IR fiber optics, e. g. hollow glass waveguides \cite{Harrington}, plastic fibers \cite{Chen}, etc. We choose the range of MIR frequencies in order to limit the thermal reservoir only up to room temperature (300K), which corresponds to one thermal photon and so satisfy the approximation of maximum one excitation ($N=1$) in the system during the evolution. The values of the coupling constants and the atom-cavity  detuning were varied in order to search for the optimal result. We mention that to satisfy the RWA we should have $2g\gg \gamma_{max}(\bar{\omega})$ \cite{Scala}. We take the values  $g_1=g_2\equiv g=J=5\gamma$, considering all the reservoirs at the same temperature and study how the atomic entanglement evolves as a function of the atom-cavity detuning, $\Delta$. The result is shown in Fig. \ref{fig2pra} from which we conclude that the atom-cavity detuning facilitates in this case the generation of a quasi-stationary atomic entanglement and for $\Delta=0.1\omega_a$ the system reaches a long-lived entanglement state. Of course, in the asymptotic limit the concurrence will vanish and the atoms eventually disentangle themselves due to the damping action of the reservoirs. 

%Fig.5 of PRA_EMO
\begin{figure} [t]
 \includegraphics[width=0.7 \textwidth]{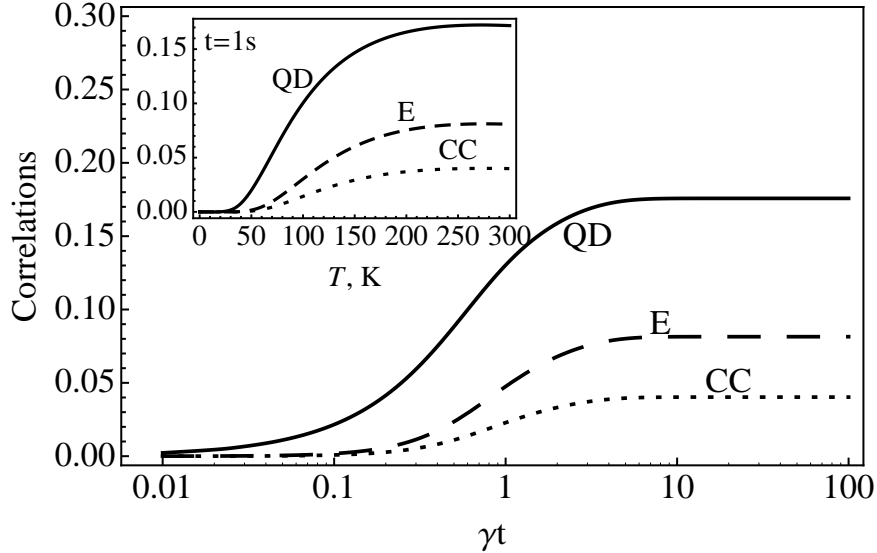}
\caption{Evolution of the quantum discord (QD), entanglement of formation (E) and classical correlations (CC) for one thermal
 excitation and the parameters chosen as in Fig. \ref{fig2pra}(c). The inset represents the same quantities as a function of the  temperatures
 of the reservoirs  calculated for a late time, $t=1$s.} 
\label{fig5pra}
\end{figure}

Nowadays researchers tend to use quantum discord rather than entanglement as a good measure of the quantum correlations \cite{Vedral, Zurek, Luo, Alber, Lu,
Fanchini}. Since in our case the two-qubit density matrix has a simplified $X$ form (see Eq. 4 in \cite{EMO}), we easily compute the quantum and classical correlations in the system by using a particular case for the algorithm discussed in \cite{Alber}, as well we checked by the approach proposed in \cite{Fanchini} and got the same results. So, in Fig. \ref{fig5pra} one finds the time evolution of the QD is very similar to that of the entanglement, but the initial growth of the discord is steeper  which implies the appearance of the quantum correlations, quantified by QD, prior to the entanglement. For a better illustration of the thermal effect under discussion, we show in the inset  the temperature dependence of the steady values (flat time plateau) of the quantum and classical correlations. 

Next, interesting results are also found in \cite{ECO}, that are related to the phenomena of sudden transitions and freezing of the classical and quantum correlations, observed and analyzed in quite different physical systems \cite{{Maziero2009}, {Auccaise2011}, {Xu}, {Pinto2013}, {He2013}, {Mazzola2010}, {Lo Franco2012}, {You2012}, {Aaronson2013}}. To search for similar effects in our system we improved considerable the approximation by increasing the number of the excitations from $N=1$ to $N=9$, and therefore we had sufficient degrees of freedom to engineer the desired initial state of the two qubits and consider many thermal excitations from the baths. For the following analysis, we consider the initial state of the two qubits (atoms) in a  Bell-diagonal (BD) state, described by an X-type density matrix defined in Eq. (\ref{rhoBD}).

\begin{figure} [t]
\centering\includegraphics[width=0.7 \textwidth]{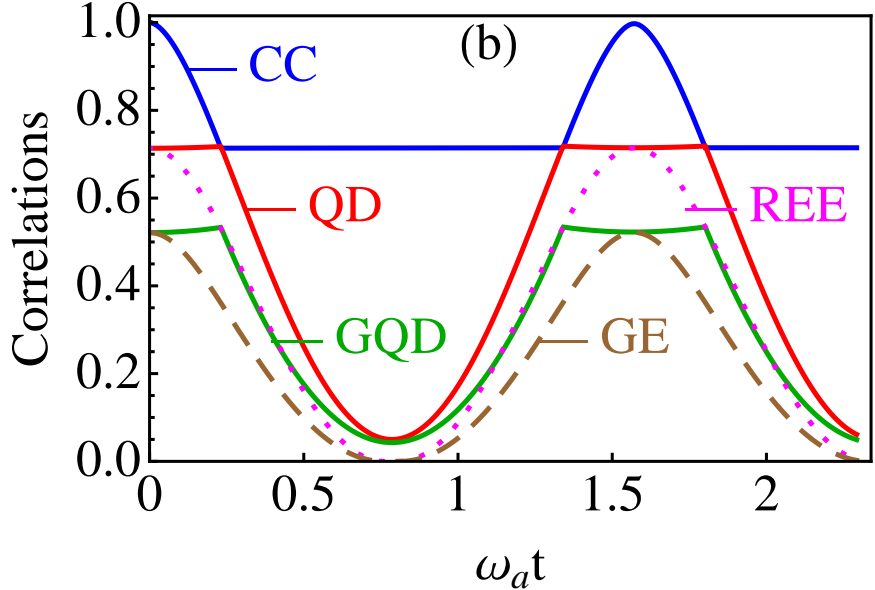}
\caption{Time evolution of the correlations: CC (blue solid),  QD (red solid), GQD (green solid), REE (magenta dotted) and GE (brown dashed) for the reservoirs at zero temperatures. The parameters are the same as in Fig. \ref{fig2ol}(a).} 
\label{fig1bol}
\end{figure}

As a result, we find, in the dynamics,  the sudden transitions between the classical and quantum correlations. In the Fig. \ref{fig1bol} we show the time evolution of the classical and quantum correlations for the case of two excitations with the qubits initially prepared in a BD state and  all the reservoirs at zero temperature. One observes the quantum-classical sudden transition in our model similar to other studied systems like \cite{{Auccaise2011}, {Xu}, {He2013}, {Mazzola2010}, {You2012}, {Lo Franco2012}, {Aaronson2013}} and others. Besides the classical correlations (CC), entropic quantum discord (QD) and relative entropy of entanglement (REE), we  also studied two geometrical measures, the geometric entanglement (GE) and geometric quantum discord (GQD) defined with Bures distance \cite{Spehner}. We evidence here that the Bures GQD and QD show similar behaviors, having flat regions not affected by the dissipation processes during a particular time period, effect known as freezing of the correlations. At the same time CC decay and meet QD in a point where a sudden change occurs. After this point the CC remains constant during another time period until other sudden change follows and so we observe periodic revival of the correlations, found in other systems too \cite{{Xu}, {Lo Franco2012}}. On the other hand, the entanglement shows a different dynamics, evidencing oscillations and no flat regions.
%Fig.1b of OL_ECO
%-------------------------------------------------------------------------------------------------------------------------------------

\begin{figure} [t]
\includegraphics[width=0.45 \textwidth]{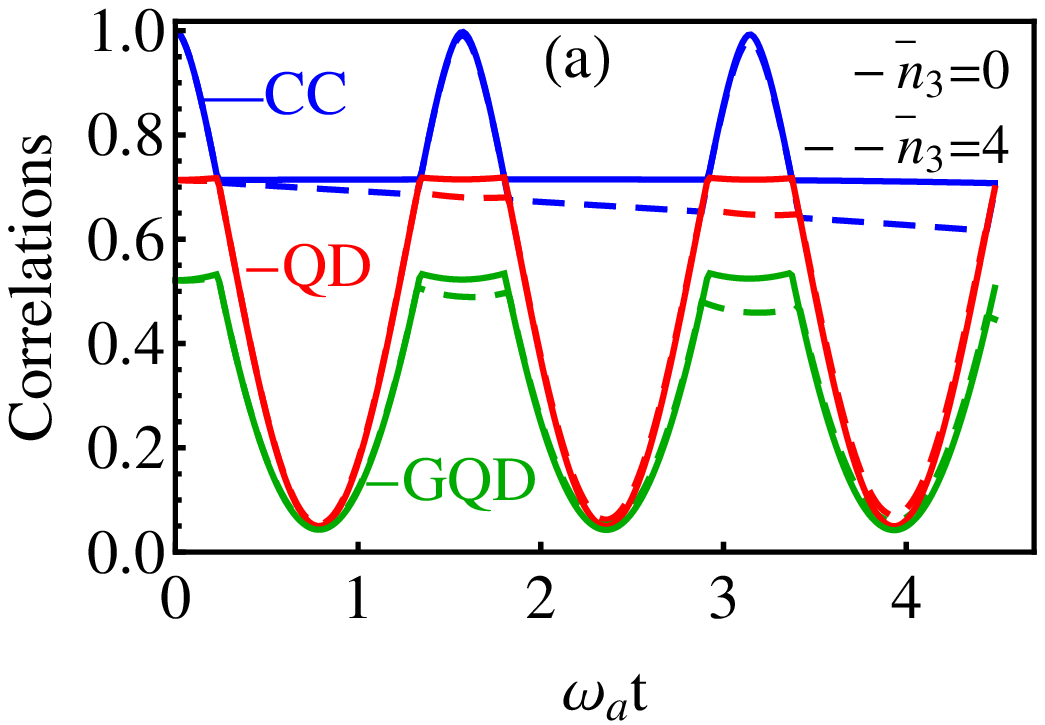}
\includegraphics[width=0.45 \textwidth]{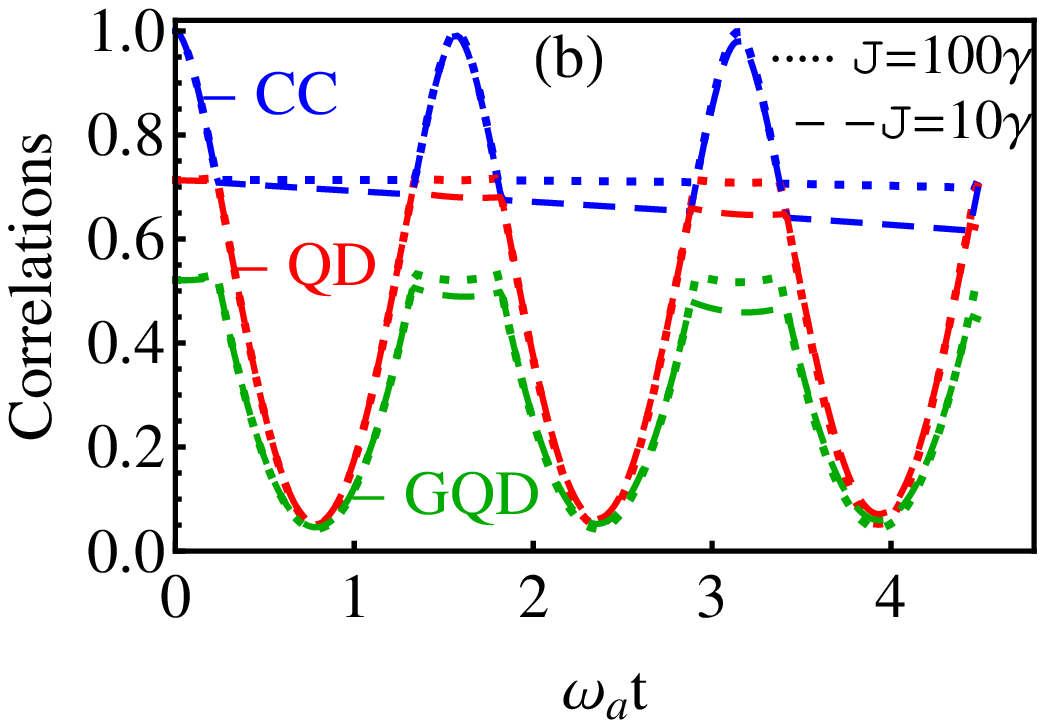}
\caption{The dynamics of the correlations: CC (blue line), QD (red line) and Bures GQD (green line) for $\gamma=0.008 \omega_a$ and (a) varying the temperature of the fiber's reservoir given by the average number of the thermal photons, i.e. $\bar{n}_3=0$ (solid line) and $\bar{n}_3=4$ (dashed line) for constant cavity-fiber coupling $J=10\gamma$; (b) varying the cavity-fiber coupling, $J=10\gamma$ (dashed line) and $J=100\gamma$ (dotted line) for constant $\bar{n}_3=4$. The initial state is defined by $\vec{c}=(1, -0.9, 0.9)$ in Eq. (\ref{rhoBD}).}
\label{fig2ol}
\end{figure}

Next, we study the dynamics of the various correlation measures for  $\omega_a/2 \pi=10$ GHz (for many experimental data see ref \cite{Haroche}). The atom-cavity couplings satisfy the constraint of the MME in a Markovian environment, i.e. $2g \gg \gamma$ and we set the values $g_1=g_2=g=10\gamma$ in the Figs. \ref{fig1bol}-\ref{fig2ol}. The values of $\gamma$ and J are tuned to evidence the effects of the thermal baths. We find that the detunings do not have an important impact on the effects of sudden transitions and freezing. We set the values $\omega_f=\omega_a$ and $\omega_a-\omega_0=0.1\omega_a$. We compute the time evolution of the atomic correlations keeping in mind the main objective of our explorations is to find the influence of the thermal baths on these correlations. In order to compute the general correlations - classical and quantum for the given system, we consider the concepts of mutual information, classical correlations and entropic quantum discord \cite{Vedral, Zurek}, as well the geometric quantum discord with Bures distance, recently developed by one of us \cite{Spehner} and independently in \cite{Aaronson2013}. 

%Fig.2 of OL_ECO

Figure \ref{fig1bol} shows the effects of sudden change and  sudden change and freezing of the correlations for the reservoirs at zero temperature .
 In our numerical analysis we find that the freezing effects of the QD and GQD decay by increasing individually or collectively the temperatures of the cavities or the fiber. In the Fig. \ref{fig2ol}(a) we show the effect of heating the fiber to four thermal photons and observe that the thermal effects act destructively on the freezing of both the entropic and geometric discords. However, the sudden transitions persist. 
 Now, could one recover from the damaging effects of the system being coupled to the thermal reservoirs? Exploring this task, we find that it is possible to engineer such a recovery by a suitable increase in the fiber-cavity coupling. Hence, in Fig. \ref{fig2ol}(b) we show that keeping the fiber's bath temperature at four thermal excitations, such recovery of the correlations is feasible if we increase the fiber-cavity coupling. Through this effect, we understand the important role of the photon as the carrier of the quantum correlations between the remote qubits in such a network.

%Fig.3 of OL_ECO
\begin{figure} [t]
\includegraphics[width=0.45 \textwidth]{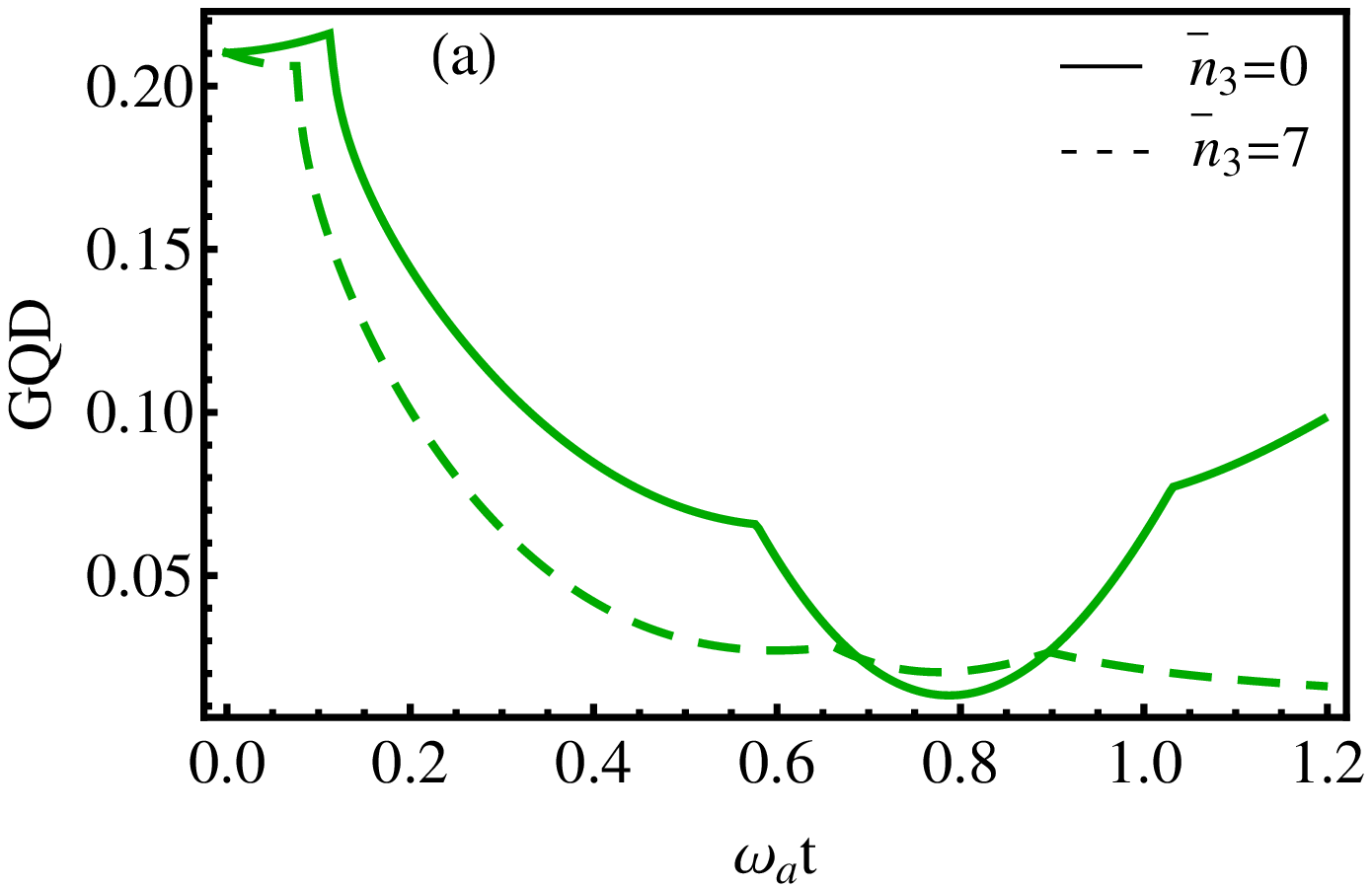}
\includegraphics[width=0.45 \textwidth]{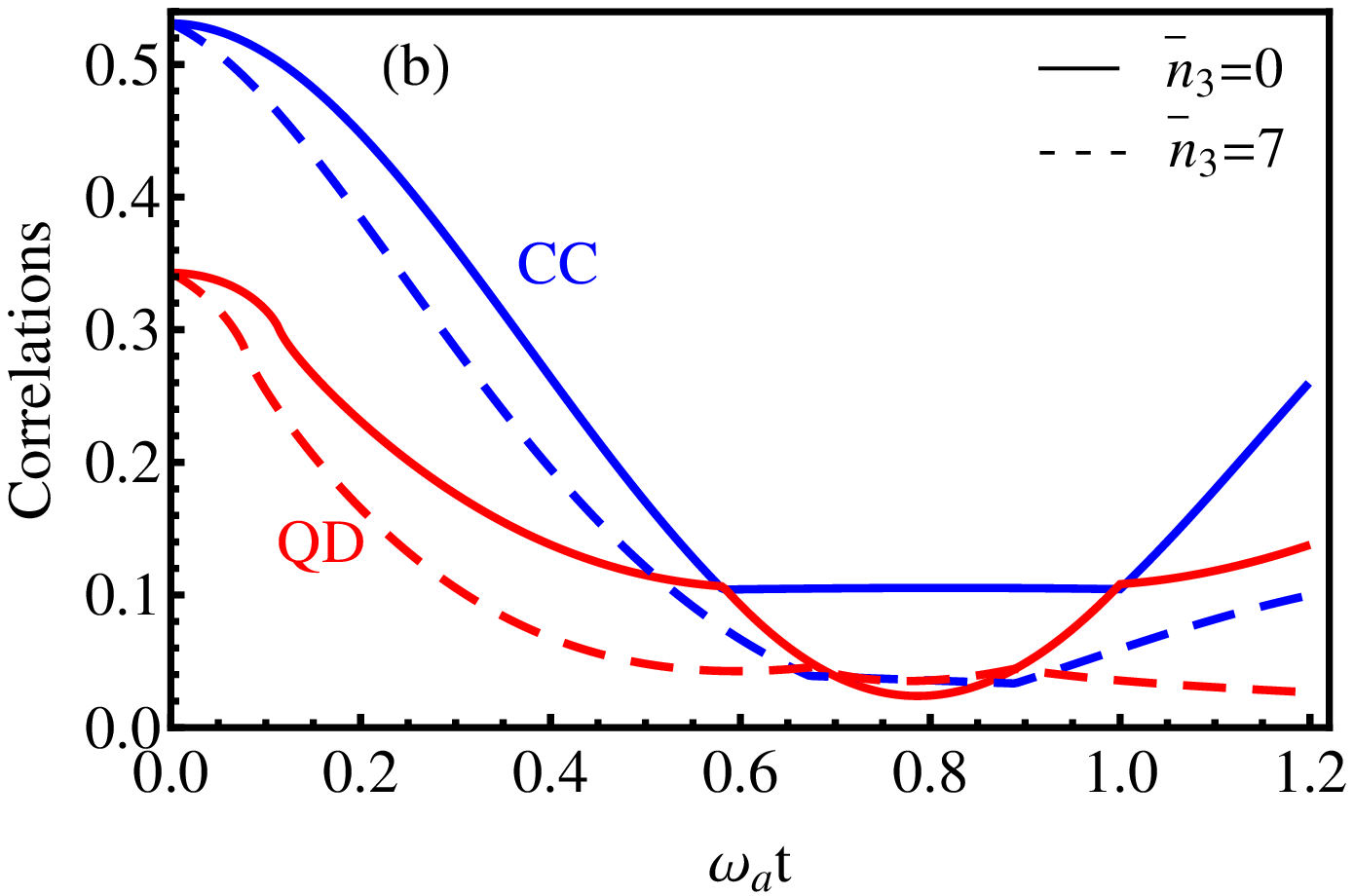}
\caption{Dynamics of the classical and quantum correlations: (a) GQD (green line) evidencing double sudden transitions, and (b) CC (blue line) and QD (red line) with only one sudden change. The parameters considered here are $\gamma=0.1 \omega_a$, $J=g=5\gamma$ for $\bar{n}_3=0$ (solid line) and $\bar{n}_3=7$ (dotted line). The initial state is given by $\vec{c}=(0.85, -0.6, 0.36)$.}
\label{fig3ol}
\end{figure}

To finish with the analysis of this model, we point out here on some different effect of multi sudden transitions in the correlations. Very recent results reported in \cite {Montealegre2013} show theoretically another interesting class of sudden transitions and freezing of the quantum correlations, which later was observed experimentally in NMR setups \cite {Paula2013}. They found the formation of an environment induced double transition of Schatten one-norm geometric quantum correlations (GQD-1), which is not observed in the classical correlations, thus a truly quantum effect. So, stimulated by this recent findings we compute the dynamics of the Bures GQD for our model and find a type of double sudden transitions somewhat different from the ones observed in \cite {{Montealegre2013}, {Paula2013}}. By this result we come to the conclusion that the both, GQD-1 and Bures GQD evidence similar quantum effects. In Fig. \ref{fig3ol} we see the double sudden changes in the dynamics of Bures GQD for the reservoirs at zero temperatures, meanwhile the CC and QD suffer only one sudden change. Next, if increase the temperature of the fiber's bath, there is a peculiar tendency to freeze the GQD and the second transition tends to disappear, at a critical temperature. Even though we donot have an adequate explanation of the physics of these effects, they still remain attractive for further fundamental and applied research. 

The experimental realization of the 2-node CQED network hinges on the possibility of realizing a quantum non-demolition (QND) measurements of the photon states in the fiber-cavities system. There is an extensive literature on QND measurements in CQED, for review see \cite {Grangier}. In our scheme we propose to measure the two-qubit density matrix under the condition that all the fields are in the vacuum state, so it is feasible to monitor the probability of this state during the temporal evolution of the system, similarly to the results presented in Fig.6 of \cite {EMO}. 

\subsection{Other types of QED networks studied in literature}

A reliable quantum network scheme that implies the transfer of information between two particles in the presence of decoherence was proposed by Pellizzari \cite{Pellizzari} more than a decade ago. 

As was shown, even when cavities and fibers are lossy, that is connected to reservoirs, one can still  transfer successfully the quantum information between locally distinct nodes (atoms) of the quantum network. The ideas of quantum network with optical fibers was developed for  arrays of cavities QED \cite{Zhang}, where the same protocol of coupling is used. The phenomenon of entanglement sudden death (ESD) was observed, where the influence of the initial state preparation of the system plays an important role. 

Another interesting model that involves the effective realization of quantum gates between two qubits in a QED network was suggested by Serafini \textit{et al} in \cite{Serafini}. The influence of the spontaneous emission and losses on the propagation of entanglement was investigated. It was observed that the coupling strengths (between the cavity mode and atoms, as well as atoms and fiber mode) can serve as the main parameters for the interaction control and distribution of quantum information.

To realize this idea in a quantum computer seems more complicated, as the number of trapped qubits should be extended and the manipulation and control between multiple particles need additional requirements. However, a many atoms interaction scheme in quantum state transfer and logical gates for two remote cavities connected via an optical fiber was investigated \cite{Yin}. In the absence of collective effects between the atoms, the coupling between the fiber and the cavity modes is sufficiently strong as compared to the atom-cavity coupling strength, and the system can be considered as two qubits resonantly coupled to a harmonic oscillator. The state transfer operation and the evolution of entangled multi-qubits are discussed in detail.            

Diamond nitrogen-vacancy centers (NV), are considered  promising candidates for qubits in the quantum applications nowadays since they have a long electronic spin coherence \cite{Togan}. A model of entanglement generation between two separated NV centers coupled to a photonic molecule has been proposed recently \cite{Liu}. A strong dependence on the hopping strength of the photonic crystal cavities and NV qubit-cavity detuning in the entanglement dynamics was remarked. Controlling with the NV the qubit-cavity coupling constants, a long time entanglement plateau was achieved. 
Several interesting schemes that are based on two NV centers coupled to a common whispering-gallery-mode microresonator (WGM) have been investigated \cite{{Wu, Schietinger}}. The dynamics of entanglement generation was studied as function of the coupling strengths of NV centers with WGM, distance between the NV centers and the state preparation of the system. The defect centers in diamond are sufficiently stable at room temperature that allow a good control with WGM interaction. Manipulating the distance between two NV centers, a maximum entanglement can be achieved.

\section{Propagation of Quantum Correlations}

The Hamiltonian of an $N$-atom-cavity system in the rotating wave approximation, joined by optical fibers, can be written as follows
\begin{equation}
\mathit{H}=\mathit{H}^{\textit{free}}+\mathit{H}^{\textit{int}},
\end{equation}
with

\begin{equation}
\mathit{H}^{\textit{free}}=\sum_{i=1}^N \omega_i^a|e\rangle_i\langle e| + \sum_{i=1}^N \omega_i^c a_i^{\dagger}a_i + \sum_{i=1}^{N-1}\omega_i^{f}b_i^{\dagger}b_i
\end{equation}

\begin{equation}
\mathit{H}^{\textit{int}}=\sum_{j=1}^N g_j(a_j^{\dagger}|g\rangle_j\langle e|+a_j|e\rangle_j\langle g|) + \sum_{j=1}^{N-1}J_j[(a_j^{\dagger}+e^{i\phi}a_{j+1}^{\dagger})b_j + H.c.],
\end{equation}
where $|g\rangle_j$ and $|e\rangle_j$ are the ground and excited states of the two-level atom with transition frequency $\omega^a$, and $a_i^{\dagger}$($a_i$) and $b_i^{\dagger}$($b_i$) are the creation(annihilation) operators of the cavity and fiber mode, respectively. The first, second and third term in $\mathit{H}^{\textit{free}}$ are the free Hamiltonian of the atom, cavity field and fiber field, respectively. In addition, the first term in the $\mathit{H}^{\textit{int}}$ describes the interaction between the cavity mode and the atom inside the cavity with the coupling strength $g_j$, and the second term is the interaction between the cavity and the fiber modes with the coupling strength $J_j$. Notice that the phase $\phi$ is due to the propagation of the field through the fiber of length $l$: $\phi=2\pi\omega l/c$ \cite{Serafini}. Also we assume that $2l\mu/2\pi c\lesssim 1$(short fiber limit), with $\mu$ being the decay rate of the cavity fields into a continuum of fiber modes.

The first two terms of $\mathit{H}^{\textit{free}}$ and the first term of $\mathit{H}^{\textit{int}}$ can be jointly diagonalized in the basis of polaritons. These states are given by,
\begin{eqnarray}
\vert n-\rangle &=& \sin(\theta)\vert n,g\rangle -\cos(\theta)\vert n-1,e\rangle\nonumber\\
\vert n+\rangle &=& \cos(\theta)\vert n,g\rangle + \sin(\theta)\vert n-1,e\rangle \nonumber\\
E_{n\pm} &=& \omega^c n+\frac{\Delta}{2}\pm \frac{\sqrt{\Delta^2 +4g^2 n}}{2}  
\end{eqnarray}
with $\Delta=\omega^a-\omega^c$, $\theta=\dfrac{1}{2}\arctan(\frac{g\sqrt{n}}{\Delta/2})$ and $n$ representing the number of photons. For simplicity we consider the resonance between atom and cavity $\omega_i^a=\omega_i^c=\omega_i$. Then, we can only have one photon per cavity, at most, because of the photon blockade, thus double or higher occupancy of the polaritonic states is prohibited \cite{blockade1,blockade2}. 

In the rotating wave-approximation and interaction picture, the hopping terms between different polaritons $\mathit{L}_{i}^{{-}^{\dagger}}\mathit{L}_{i+1}^{+}$ and $\mathit{L}_{i}^{{+}^{\dagger}}\mathit{L}_{i+1}^{-}$, with $\mathit{L}^{{\pm}^{\dagger}}=\vert 1\pm\rangle\langle 0\vert$ defined as the creation operator, are fast rotating and they average zero. This implies that if we started creating a polariton with $\mathit{L}^{{-}^{\dagger}}$, the state $\vert 1 +\rangle$ will never show up. Finally, we restrict the subsystem to only two states, $\vert G\rangle=\vert 0\rangle$(ground state) and $\vert E\rangle =\vert 1 -\rangle =\frac{1}{\sqrt{2}}(\vert 1 g \rangle-\vert 0 e \rangle)$(excited state), and from now on we will omit label ``$-$" on $\mathit{L}$.

\begin{equation}
\mathit{H}=\sum_{i=1}^N(\omega_i-g_i)|E\rangle_i\langle E|+ \sum_{i=1}^{N-1}\omega_i^{f}b_i^{\dagger}b_i+\sum_{i=1}^{N-1}\frac{J_i}{\sqrt{2}}[(L_i^{\dagger}+L_{i+1}^{\dagger})b_i + (L_i+L_{i+1})b_i^{\dagger}]
\end{equation}

In the present model we are not interested in the fiber, so we want to eliminate it. One way is solving the complete Hamiltonian and finally trace or measure with a projecting operator on the fiber \cite{Mont, EMO}. Another way is to eliminate the fiber first, and then analyze the time evolution of the system. We propose to discuss two different approaches in the latter case.

On the one hand, we can eliminate adiabatically the fiber mode and obtain the effective Hamiltonian \cite{mancini}
\begin{equation}\label{hamilt_a}
\mathit{H}_{\textit{a}}^{\textit{eff}}= \sum_{i=1}^{N}\omega_{i}^{\prime}\mathit{L}_{i}^{\dagger}\mathit{L}_{i} + \sum_{i=1}^{N-1}J_i^{\prime}(\mathit{L}_{i+1}^{\dagger}\mathit{L}_{i} + \mathit{L}_{i}^{\dagger}\mathit{L}_{i+1}),
\end{equation} 
where $\omega_{i}^{\prime}=\omega_i -g_i-\frac{2J_i^2}{\omega_i^f} + \frac{J_i^2}{\omega_i^f}\delta_{i,1} + \frac{J_i^2}{\omega_i^f}\delta_{i,N}$ and $J_i^{\prime}=-\frac{J_i^2}{\omega_i^f}$.

On the other hand, we can use perturbation theory to eliminate the fiber \cite{cohen,raul1}.  We assume first that all cavities and fibers, and their corresponding coupling constants are equal, i.e., $\omega_i=\omega$, $\omega_i^f=\omega^f$, $g_i=g$ and $J_i=J$, and second, that the total detuning $\delta=(\omega-g)-\omega^f\gg J$. Then, we project the fiber state into the zero photon mode, generating an effective Hamiltonian given by

\begin{equation}\label{hamilt_p}
\mathit{H}_{\textit{p}}^{\textit{eff}}= \sum_{i=1}^{N-1}\lambda (|E\rangle_i\langle E| + |E\rangle _{i+1}\langle E|) + \sum_{i=1}^{N-1}\lambda (\mathit{L}_i^{\dagger}\mathit{L}_{i+1} + \mathit{L}_{i+1}^{\dagger}\mathit{L}_i),
\end{equation}
where $\lambda=\frac{J^2}{2\delta}$. For this approach it is important that the fibers be weakly coupled to the cavities. Also, when obtaining this effective Hamiltonian we allow just one excitation in the chain.

A different model, still having an optical fiber, was proposed by Zhong \cite{zhong}. In this work, he used a configuration where one cavity (e.g. the central cavity) is connected through optical fibers to several cavities, which are not connected between  them.
In this proposal, it is possible to generate entangled states for multiple atoms trapped in distant cavities, connected by optical fibers. There is also considered resonant interaction between atoms, cavities and fibers, so the interaction time is short, which is an important factor when dealing with decoherence.
 
\begin{figure}[ht]
\centering
\includegraphics[width=8.3cm]{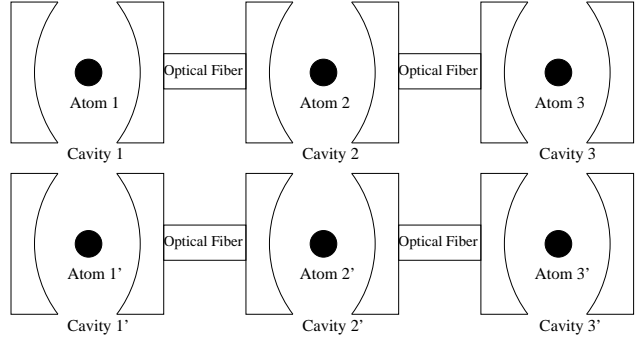}
\caption{Array of two rows of three cavity-atom systems.}
\label{fig2}
\end{figure}

A straight generalization of the one chain model was studied by Zhang \textit{et al.} \cite{Zhang}, proposing a system with two non-interacting chains, see Fig. \ref{fig2}. This idea is quite interesting, since it opens a new series of applications, different from the one chain models.
However, this model, does not include losses and it is limited to the analysis of the propagation of one kind of quantum correlation (entanglement). In an endeavour to improve the model, we added losses and did a more exhaustive study \cite{raul1}, including the propagation of entanglement and quantum discord, the distribution of the entanglement, and we discussed a possible application for quantum communications \cite{raul2}.  

We are interested in sending information through both chains. This implies that starting with two correlated qubits corresponding, for example, to the $11'$ cavities. We then study the dynamics of our system such that after some time, the $33'$ pair becomes correlated. We used the following notation, $|\Psi\rangle=|X_1X_{1'}X_2X_{2'}X_3X_{3'}\rangle$, where $X$ could be $G$ or $E$.
 The time evolution of the whole system is given by the Hamiltonian $\mathit{H}=\mathit{H}_1\otimes\mathit{H}_2$, with $\mathit{H}_{\lbrace 1,2\rbrace}$ defined in Eq.(\ref{hamilt_p}). We took the parameters: $J=2\pi\cdot30\,GHz$, $\delta=2\pi\cdot300\,GHz$ and studied the dynamics for two different initial states:

\begin{eqnarray}\label{initialstate1}
&|\Psi\rangle_a=\sin(\theta)|GEGGGG\rangle + \cos(\theta)|EGGGGG\rangle\nonumber\\
&|\Psi\rangle_b=\sin(\theta)|GGGGGG\rangle + \cos(\theta)|EEGGGG\rangle
\end{eqnarray}

\begin{figure}[ht]
\centering
\includegraphics[width=0.7 \textwidth]{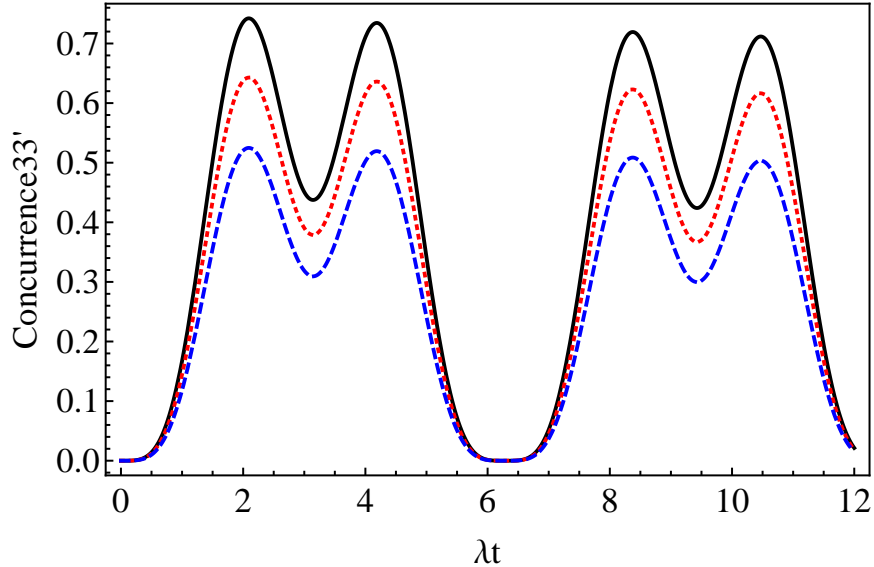}
\caption{Concurrence, C in case of the cavities $33'$ for the initial condition $|\Psi\rangle_a$ with constant $\gamma=0.01$ and varying the angle as: $\theta =\pi/4$ (solid line), $\theta =\pi/3$ (red-dotted) and $\theta =\pi/8$ (blue-dashed).}
\label{fig3}
\end{figure}

We found that the transmission properties of the entanglement depend strongly on the initial conditions. For example, we observed that for the initial state $|\Psi\rangle_a$, $74.2\%$ of the concurrence in the cavity-pair $11'$ is transmitted to the $33'$ pair, independent of the angle $\theta$. On the other hand, for $|\Psi\rangle_b$ the transmission depends strongly on $\theta$. For example, for $\theta=\pi/3$ we get $63\%$ and for $\theta=\pi/8$, $28\%$. The final concurrence $33'$ is shown in Fig.\ref{fig3}, for the initial state $|\Psi\rangle_a$.

Now, we  compare the quantum discord and the entanglement to enquire which of this quantifiers is more robust against decoherence in this system. In some of the following calculations, when comparing the various measures of correlations,  it will be  more convenient to calculate the entanglement of formation ($\mathit{E}$), rather than the concurrence. Next, we plot both, for the cavity $33'$, and study the time evolution of the system.

\begin{figure}[ht]
\centering
\includegraphics[width=0.45 \textwidth]{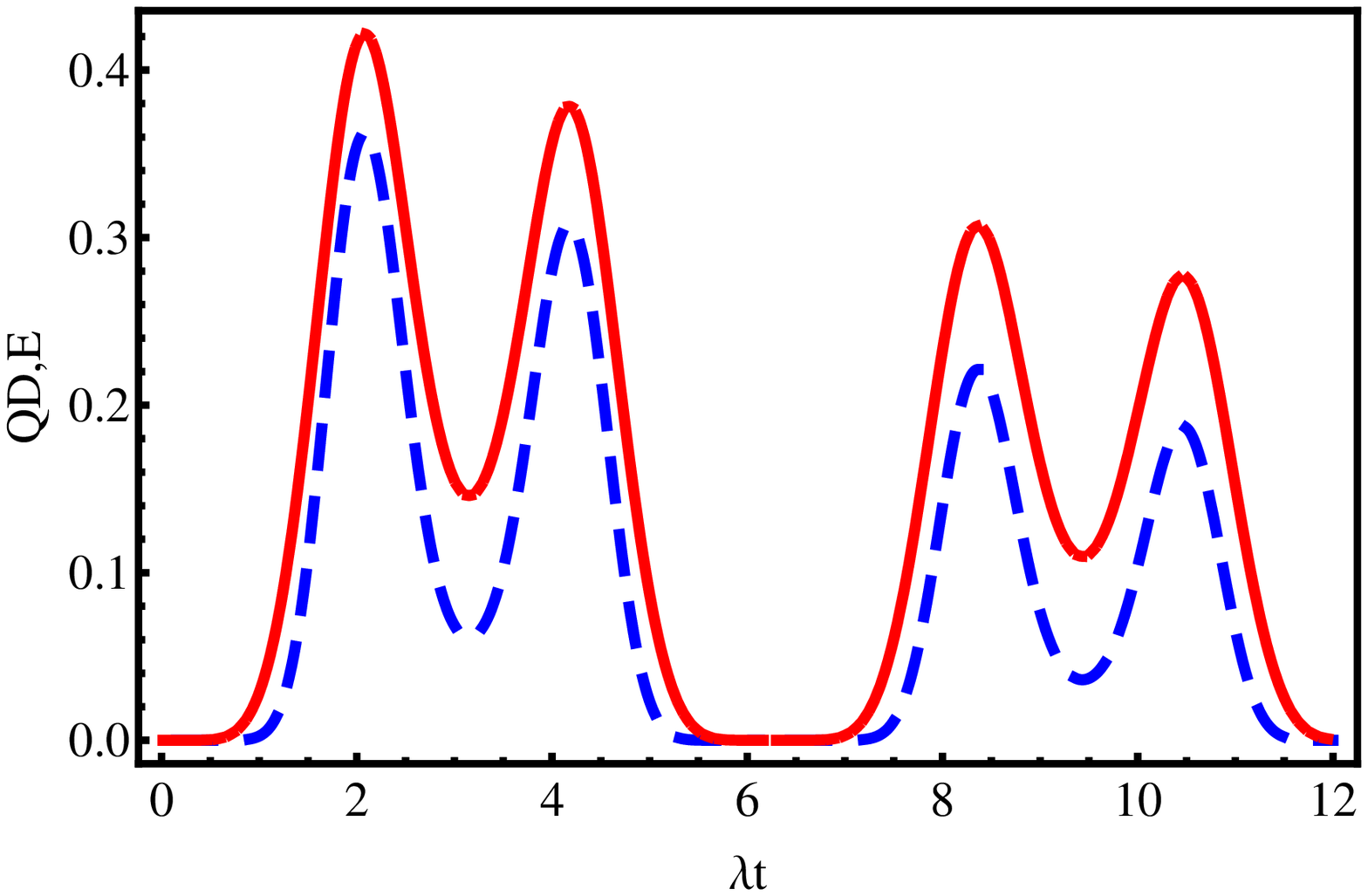}
\includegraphics[width=0.45 \textwidth]{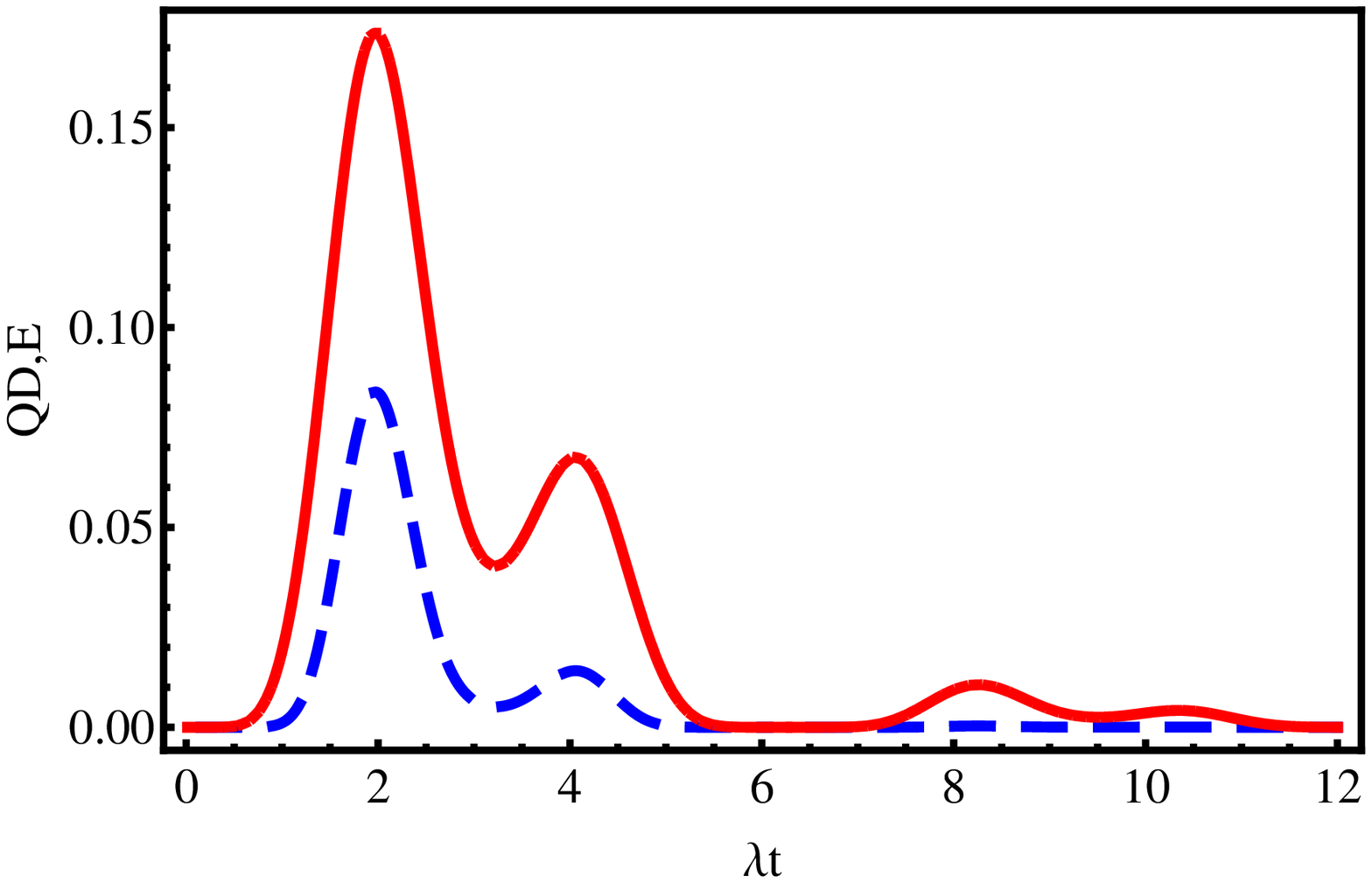}
\caption{Quantum discord, QD (red solid) is more robust than entanglement of formation, E (blue-dashed) in case of the cavities $33'$ for the initial condition $|\Psi\rangle_b$ with constant $\theta=\pi/4$ and varying $\gamma$ as:  $\gamma=0.05$ (left panel) and $\gamma=0.5$ (right panel).}
\label{fig4}
\end{figure}

From Fig.\ref{fig4} it can be seen that the quantum discord, remains bigger than the entanglement of formation. Notice that near the origin, QD grows up before than E, being as a precursor of the generation of quantum correlations. Asymptotically for all $\gamma$, QD tends to be larger than E, even when the latter vanishes, which is in agreement with previous work in cavity QED \cite{bellomo}.

As another example, we extend the comparison changing the previous pure initial condition for a Werner state(W) for the cavities $11'$, with the rest of the cavities being in the ground state($|G\rangle$) 
 
\begin{equation}\label{initialstate2}
\rho (0)=(\dfrac{1-a}{4}\mathit{I}+a|\psi\rangle\langle\psi|)\otimes|G_2G_{2'}G_3G_{3'}\rangle\langle G_2G_{2'}G_3G_{3'}|
\end{equation}

where $\mathit{I}$ is the identity operator for two qubits and $|\psi\rangle =\frac{1}{\sqrt{2}}(|E_1G_{1'}\rangle +|G_1E_{1'}\rangle)$.

\begin{figure}[ht]
\centering
\includegraphics[width=0.45 \textwidth]{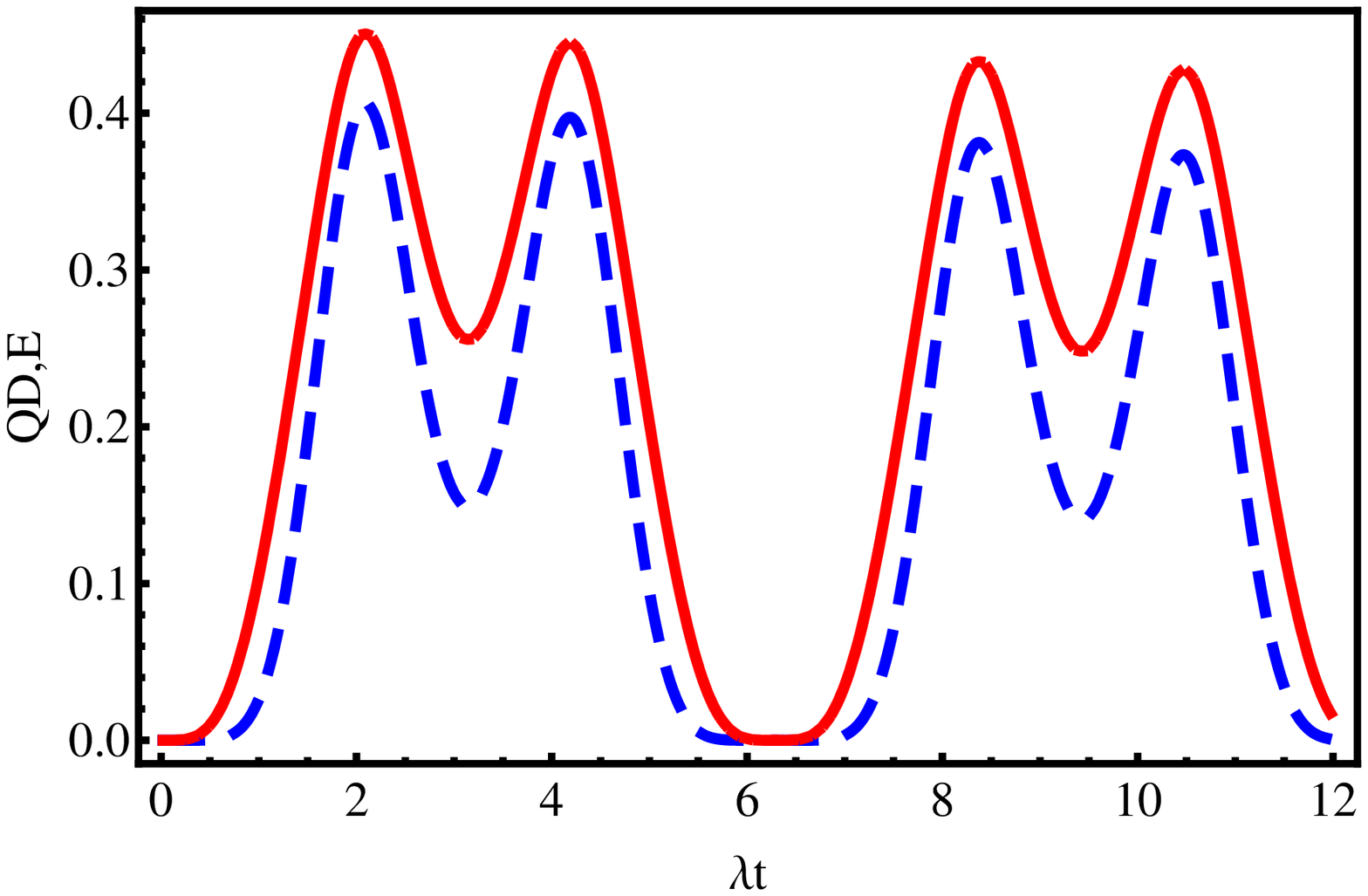}
\includegraphics[width=0.45 \textwidth]{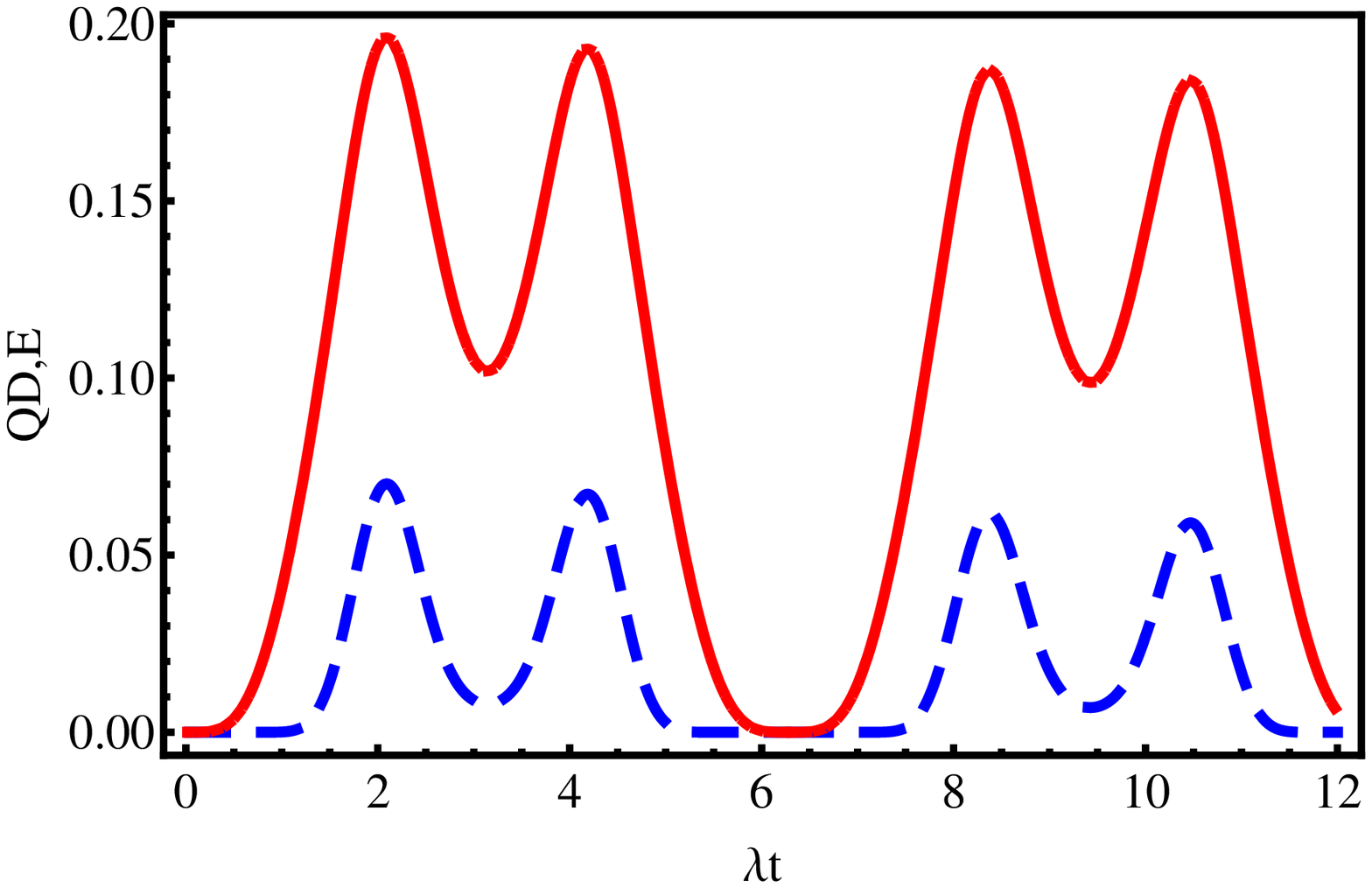}
\caption{Quantum discord, QD (red solid) and entanglement of formation, E (blue-dashed) for the cavities $33'$ with Werner initial state, where $a=0.9$ (left panel) and $a=0.6$ (right panel).}
\label{fig5}
\end{figure}

\begin{figure}[ht]
\centering
\includegraphics[width=0.45 \textwidth]{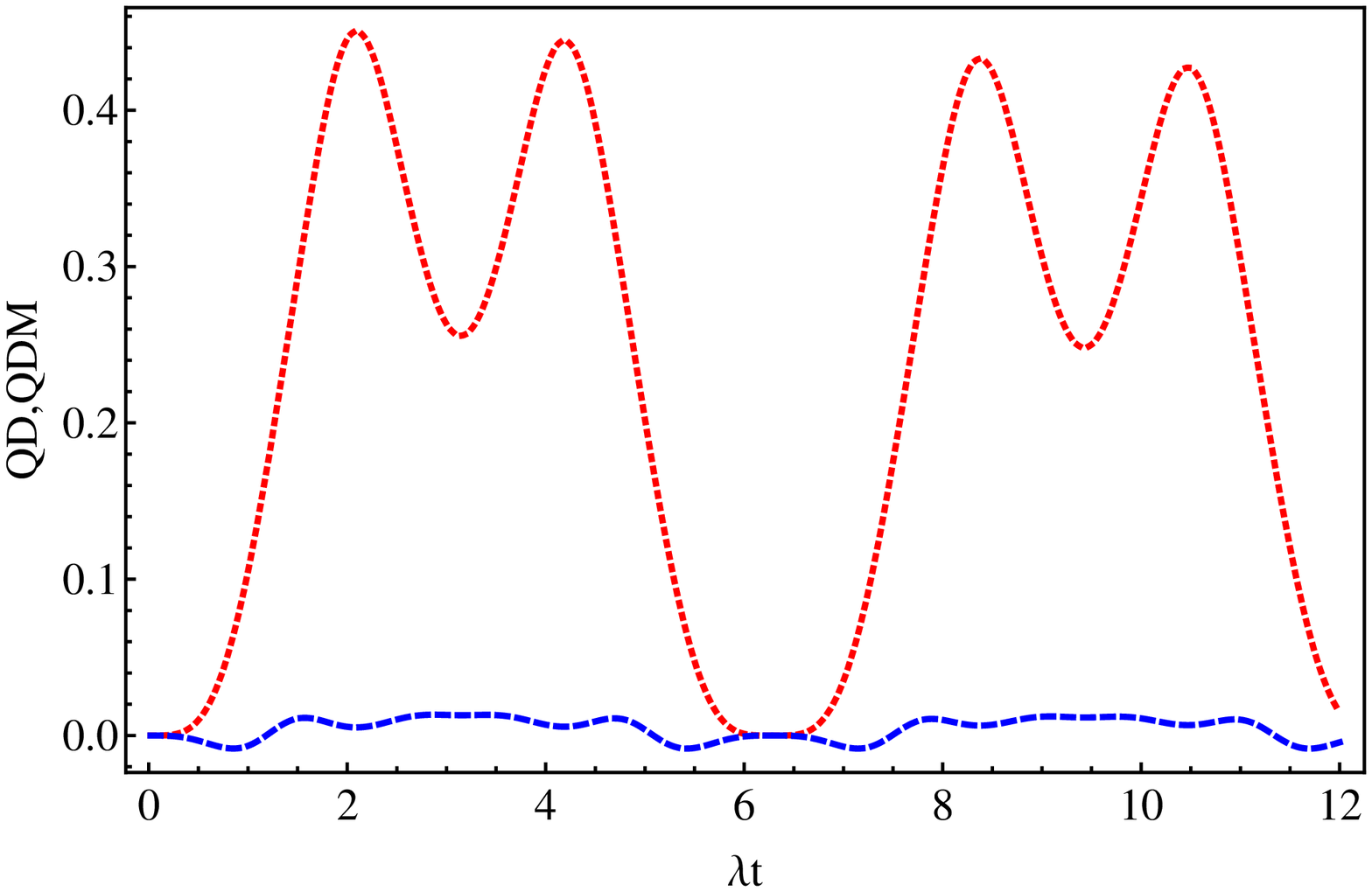}
\includegraphics[width=0.45 \textwidth]{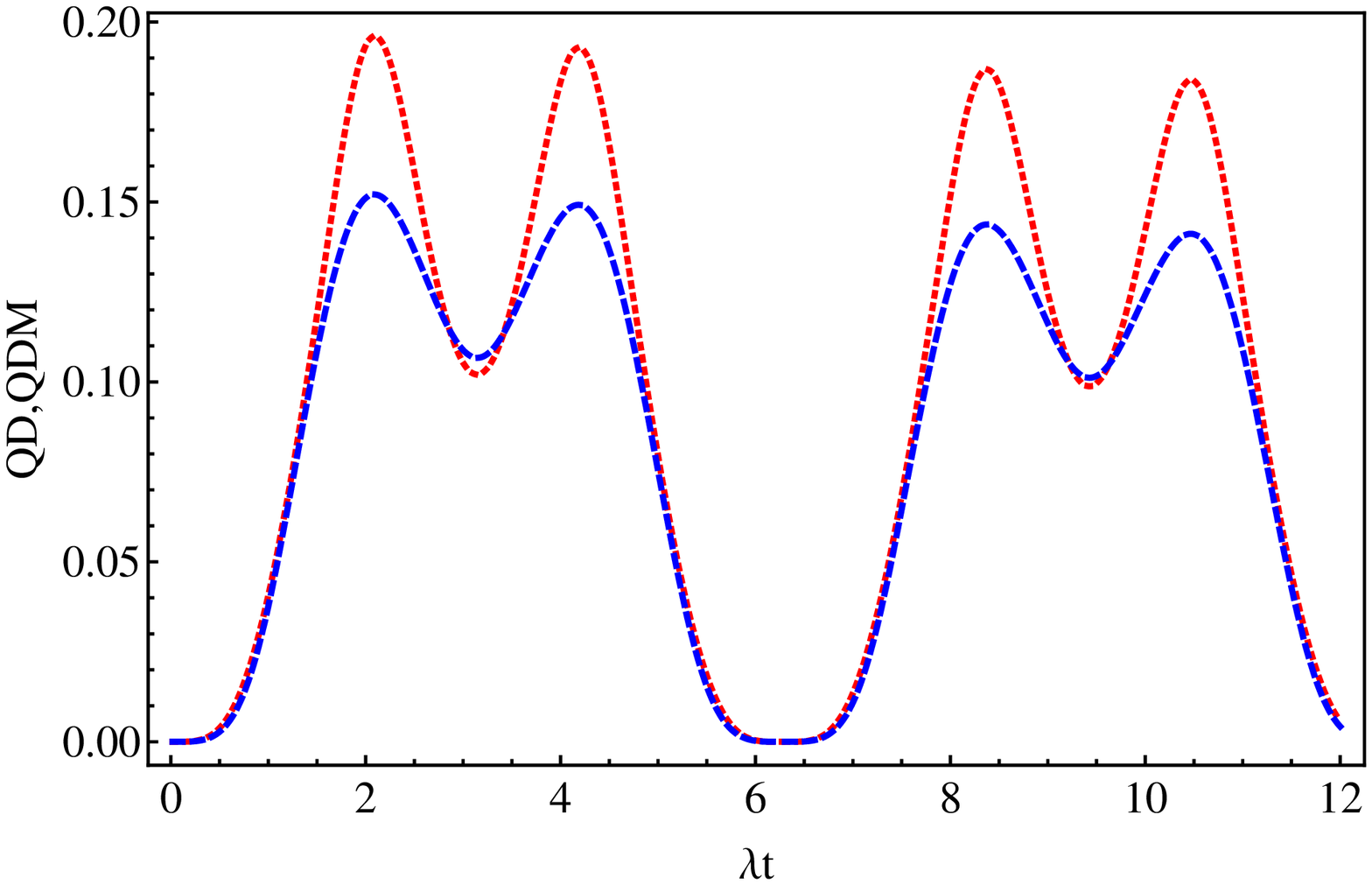}
\caption{Quantum discord after the measure, QDM (blue-dashed) varies considerably for highly pure states, as compared with quantum discord, QD (red-dotted) for the cavities $33'$ considering: $a=0.9$ (left panel) and $a=0.6$ (right panel).}
\label{fig6}
\end{figure}

In figures (\ref{fig5}), we plot the $QD$ and $E$, and see the behavior arising from varying the mixedness of the initial state. When the system is nearly a pure state ($a=0.9$), there is not a big difference between the two curves (Fig.$\ref{fig6}$). However, when the system becomes more mixed ($a=0.6$), there is a substantial difference between $\mathit{QD}$ and $\mathit{E}$ and obviously the $\mathit{QD}$ is the better option for the propagation of the information (Fig.$\ref{fig5}$).

Next, we analyze the effect on QD of $33'$ if a third person (an eavesdropper) performs a measurement on cavity $2$ for example. Let's choose a projective measure such as $\Pi=|G_2\rangle\langle G_2|$. The main idea behind this, is to compute the QD after the measurement($\mathit{QDM}$) at the $33'$ pair and compare it with the undisturbed case. From Fig.$\ref{fig6}$ left, we can see that for a nearly pure maximally entangled state, the curve corresponding to the Quantum Discord after the measurement is reduced to almost zero (blue-dashed) as compared with  the undisturbed Quantum Discord without any measurement (red-dotted).
It is quite apparent that in this case we have a very good instrument to detect any external measurement. However if the state becomes more mixed ($a=0.6$), the discrimination becomes inconclusive, since in Fig.$\ref{fig6}$ right, we do not observe relevant differences anymore between the two curves.

\section{Distribution of Quantum Correlations}

As we discussed in the Introduction, Coffman et al \cite{wootters} consider "`distributed entanglement"' or the fact that quantum correlations cannot be freely shared among several particles. They go even further and define the "`tangle"' as a measure to describe the multipartite (beyond the bipartite) entanglement.

In particular, in a three particle system, $\tau_{123}$ represents the collective entanglement of the three qubits, or the "three-qubit entanglement".
 
Unfortunately, these concepts are only valid for pure states.
For mixed states, which is the more realistic case of open systems, governed by Master Equations, if the losses are not too large, we can only estimate the lower and upper bounds of the tangle.

In the present section, we study the tangle for pure states as well as the upper and lower bounds for mixed states, under various initial conditions.
We take  relatively moderate losses, in order to have reasonable good approximate bounds.

\begin{figure}[ht]
\centering
\includegraphics[width=0.75 \textwidth]{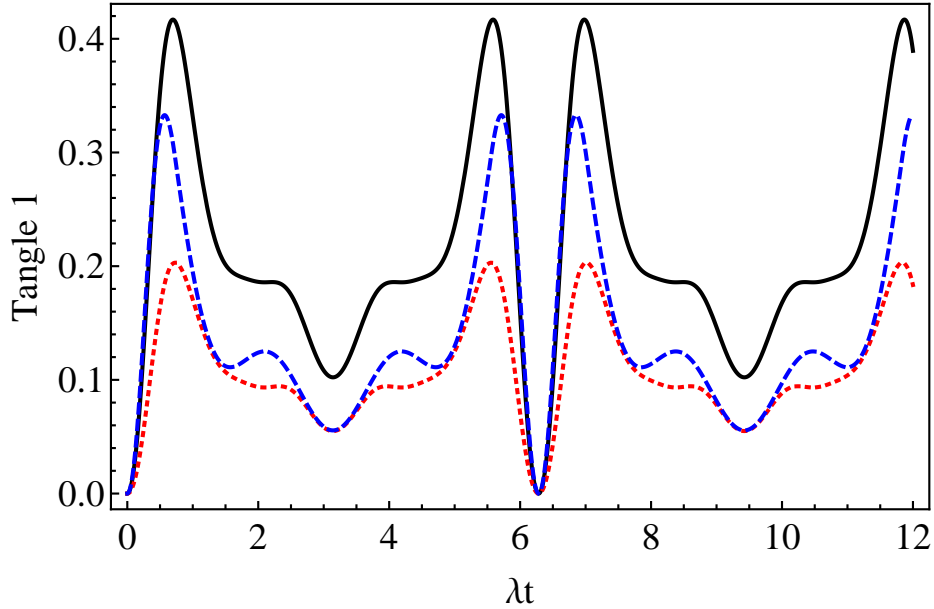}
\caption{Tangle for the initial condition $|\Psi\rangle_b$ with $\gamma=0$ for different angles: $\theta =\pi/4$ (black solid), $\theta =\pi/3$ (red-dotted), $\theta =\pi/8$ (blue-dashed) }
\label{fig7}
\end{figure}

Going back to the previous models, for one chain, with the system evolving with the Hamiltonian defined in Eq.(\ref{hamilt_p}), no tangle was found. A possible reason is because of the restriction of only one excitation in the chain, since the tangle is a collective effect and needs more than one excitation in the system. We can see this more clearly in the two chain model. Here, for the initial state $|\Psi_a\rangle$ in Eq.(\ref{initialstate1}), the tangle is always zero, independent of $\theta$. On the other hand, the state $|\Psi_b\rangle$ has non vanishing tangle, see Fig. \ref{fig7}. Then, the presence of two excitation in the last case is responsible for the multipartite correlation. 

In figure(\ref{fig7}) we plotted the tangle, taking cavity $1$ as the reference one, $\tau=\mathit{C}^2_{1(23...)}-\mathit{C}_{12}^2-\mathit{C}_{13}^2-\mathit{C}_{11^{\prime}}^2 -\mathit{C}_{12^{\prime}}^2-\mathit{C}_{13^{\prime}}^2$. We set the cavity losses to zero ,such that the whole system remains pure and Eq.(\ref{tangle}) is correct. In all cases, the initial tangle is zero, since we start with a bi-partite entanglement between the first pair of cavities and therefore there are no higher-order correlations.

\begin{figure}[ht]
\centering
\includegraphics[width=0.75 \textwidth]{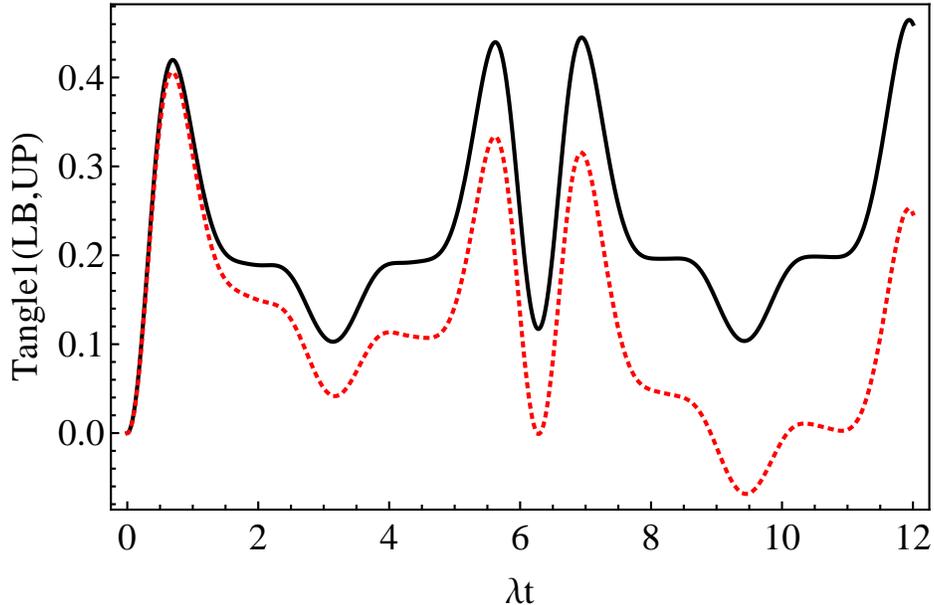}
\caption{Tangle for the initial condition $|\Psi\rangle_b$ with $\theta =\pi/4$, $\gamma=0.01$ and two different bounds: upper bound, UB (solid) and lower bound, LB (red-dotted);}
\label{fig8}
\end{figure}

If we now turn on the interaction with the individual reservoirs, the situation becomes more involved, and in principle it would require a complex convex roof optimization procedure. Nevertheless, when the system experiences losses, if these are moderate, we can still estimate lower and upper bounds to the tangle, in the case where the mixedness of the system, measured through $tr[\rho^2]$, varies slowly between $1$ and $0.89$ for $\gamma=0.01$. For higher losses, like $\gamma=0.5$, the gap between the bounds is significantly bigger and the above approximations fail. In Fig. \ref{fig8} we observe the upper and lower bounds of the tangle. In the lower bound approximation, we need to guarantee that the system is weakly mixed and strongly entangled. In particular for $\lambda t\approx 9$, the lower bound becomes negative. On the other hand, we notice that in this region, $\mathit{C}_{1(23...)}^2$ is comparatively small, thus violating the assumptions made by the lower bound approximation, and therefore the results are unreliable. However, for $\lambda t\in \lbrace 0,6 \rbrace $, the area between the upper and lower bound is rather small, giving us a good estimation of the tangle.

Finally, if the system is used as a channel, our best option is to use states like $|\Psi_a\rangle$ as the initial condition, since the distribution or multipartite entanglement vanishes, finding only bi-partite quantum correlations and as a consequence we get higher values of the concurrence at the extreme of the chains. For low losses, the entanglement of formation is a good measure of the quantum correlations. However, as previously shown, the quantum discord is more robust against decoherence, thus is a better measure for higher loss rates. If our purpose is to distribute the quantum correlations among the various elements of our system, we choose $|\Psi_b\rangle$ as our initial condition, since we have a considerable multipartite entanglement or tangle. Furthermore, we observed from Fig. \ref{fig7} that the tangle deteriorates rapidly, as we depart from the Bell states ($\theta=\pi/4$).

\section{Summary and Discussion}
In this Review we have dealt with various aspects of the quantum correlations using different measures such as the Entanglement of Formation, Concurrence, Quantum Discord, Relative Entropy of Entanglement and Geometric Quantum Discord with Bures Distance.
The studies we have covered are related to the phenomena as generation, propagation, distribution, thermal and critical effects of these correlations in the models of cavity QED networks.
We have discussed the possibility of generating atomic entanglement with atoms located at distant cavities and connected via an optical fiber, and finding a wide time plateau of the concurrence between the atoms, even when the system is connected to various reservoirs, implying lossy cavities and fiber.
Dissipation and thermal effects are normally considered destructive from the quantum correlations viewpoint. However, examples are shown that under certain conditions, these effects may contribute to the generation of these correlations. These effects are found in cavity QED networks as well as in other physical systems. 
The way the quantum correlations propagate and distribute themselves between various components of a given physical system is still an open problem.
We explore analytically and numerically the propagation and distribution of quantum correlations through two chains of atoms inside cavities joined by optical fibers. This particular system can be used as a channel of quantum communication or a network of quantum computation. One can readily select the appropriate initial condition in order to optimize the performance for the former or latter application.
Finally, we discuss the thermal effects on the sudden changes and freezing of the classical and quantum correlations in a cavity quantum electrodynamic network with losses. For certain initial conditions, double transitions in the Bures Geometrical discord are found. One of these transitions tend to disappear at a critical temperature, hence freezing the discord.

We acknowledge the financial support of the Fondecyt projects no.100039 and no.1140994, the project Conicyt-PIA Anillo ACT-112, "Red de analisis estocastico y aplicaciones", as well Pontificia Universidad Catolica de Chile and Conicyt doctoral fellowships.

\end{document}